\documentclass[%
 reprint,
superscriptaddress,
longbibliography,
 amsmath,amssymb,
 prl,
 citeautoscript,
]{revtex4-1}

\usepackage{graphicx}
\usepackage{dcolumn}
\usepackage{bm}
\usepackage{amsmath}
\usepackage{amssymb}
\usepackage{amsthm}
\usepackage{units}
\usepackage{gensymb}
\usepackage{xcolor}

\begin{document}

\title{Emergent Kelvin waves in chiral active matter}

\author{Anthony R. Poggioli}
\email{arpoggioli@berkeley.edu}
\affiliation{Department of Chemistry, University of California, Berkeley}
\affiliation{Kavli Energy NanoScience Institute, Berkeley, California}

\author{David T. Limmer}
\email{dlimmer@berkeley.edu}
\affiliation{Department of Chemistry, University of California, Berkeley}
\affiliation{Kavli Energy NanoScience Institute, Berkeley, California}
\affiliation{Materials Science Division, Lawrence Berkeley National Laboratory}
\affiliation{Chemical Science Division, Lawrence Berkeley National Laboratory}

\date{\today}

\maketitle


\noindent {\bf The phenomenological equations of hydrodynamics describe emergent behavior in many body systems. 
Their forms and the associated phenomena are well established when the quiescent state of the system is one of thermodynamic equilibrium, yet away from equilibrium relatively little is firmly established. Here, we deduce directly from first principles the hydrodynamic equations for a system far from equilibrium, a chiral active fluid in which both parity and time-reversal symmetries are broken. With our theory, we rationalize the emergence of a spontaneous boundary current in the confined fluid, a feature forbidden at equilibrium, which allows us to extract estimates of transport coefficients that we favorably compare to forced flows. The hydrodynamic solution reveals that the boundary current is analogous to a quasigeostrophic coastal current, a well known phenomenon in oceanography. Such currents are conjugate to a class of chiral waves called Kelvin waves. Motivated by this analogy, we demonstrate that an acoustic chiral Kelvin wave mode also exists in confined chiral active matter in the absence of an imposed rotation, originating from the spontaneous emergence of a Coriolis-like parameter in the bound modes of a chiral fluid.}

Chiral active fluids are composed of actively rotating constituents, and this activity breaks both time-reversal and parity symmetries. This leads to the emergence of an exotic transport coefficient called the Hall or odd viscosity, $\eta_O$ \cite{Avron}. Qualitatively, $\eta_O$ has the effect of linking forcing to orthogonal fluid motion \cite{cargo, Hall-like}, and is therefore phenomenologically similar to the Coriolis parameter that appears in oceanic and atmospheric flows \cite{Gill, science_topology, gauge_theory}. We show that this analogy becomes exact when considering the boundary-trapped collective motions in a chiral fluid, and we exploit this analogy to demonstrate the existence of a chiral, boundary-trapped acoustic wave mode, analogous to oceanographic coastally bound Kelvin waves.

\begin{figure*}[t]
	\centerline{\includegraphics[width=17cm]{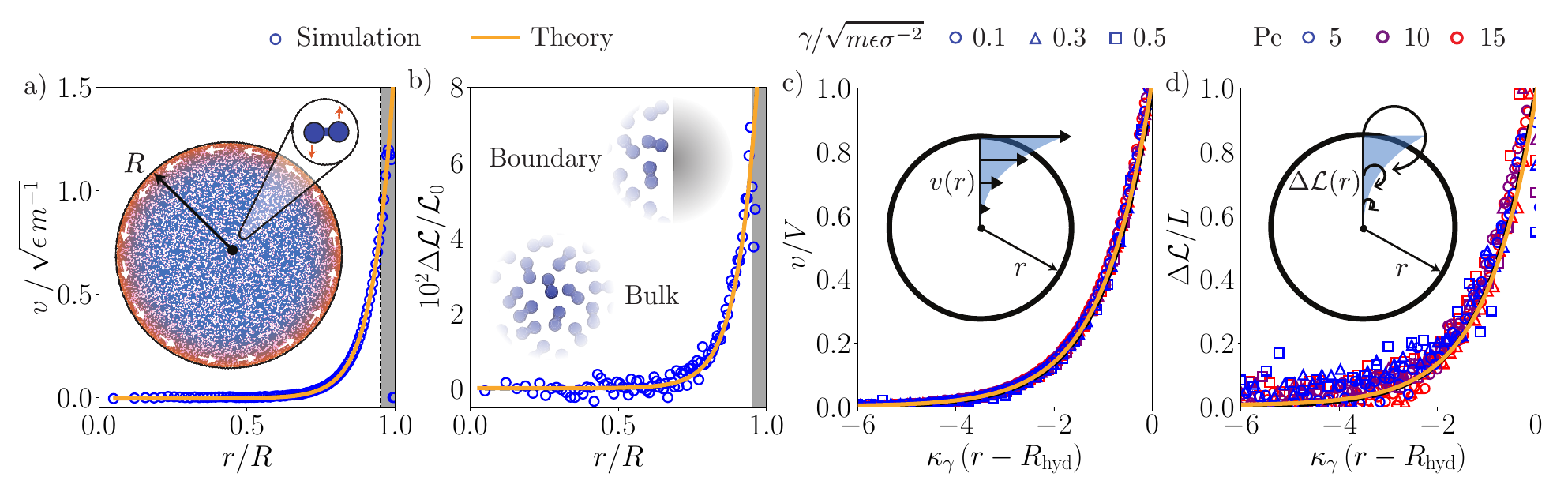}}
    \caption{Structure of spontaneous boundary current in a chiral active fluid. a), Velocity profile obtained for ${\rm Pe} = 5$, $\gamma = 0.1 \, \sqrt{m \varepsilon \sigma^{-2}}$. b), Spin anomaly profile for same parameters. c) Velocity and d), spin anomaly profiles collapsed according to theoretical prediction for profile structure. Inset in a is a system snapshot. Gray shaded regions in a and b indicate region beyond hydrodynamic radius. Solid lines are theoretical predictions from Eq. \ref{eqn:v_BC} and \ref{eqn:DL_BC}}
\label{fig:Fig1}
\end{figure*}

In addition to the odd viscosity, broken parity and time-reversal symmetries also lead to the spontaneous formation of a boundary current in confined chiral fluids, with its propagation direction determined by the chirality of the system-- or equivalently, by the sign of the odd viscosity \cite{waves_Kranthi, free_surface, lubensky_etaR6, crystal}. This spontaneous collective current is a consequence of broken detailed balance, characterized by the emergence of microscopic currents, along with the global rotational symmetry of the system, which requires the organization of microscopic currents into a collective, angular momentum carrying, boundary current. Previous studies have treated the boundary current phenomenologically, suggesting that its formation is due to a hypothetical dissipative coupling between the fluid vorticity and rotor frequency, along with the frustration of free spin at the particle boundary \cite{lubensky_etaR6, waves_Kranthi,free_surface,crystal}. This frustration leads to a local suppression of the rotor frequency, and, nominally, to a compensating increase in the fluid vorticity. However, inhomogeneities in hydrodynamic densities within a few particle diameters of an interface are expected in all hydrodynamic systems, and the frustration of spin in this microscopic contact layer need not have any influence on the hydrodynamic structure of the boundary current beyond this layer. We show that the boundary current forms in the absence of spin-vorticity coupling, deduced directly from a first principles evaluation of the emergent hydrodynamics of the system. Its formation is most clearly understood as a consequence of angular momentum conservation (SI). Injection of angular momentum through particle activity is balanced globally by substrate friction. However, frustration of free spin in the system at finite density due to inter-particle interactions leads to a reduction of the total spin and therefore to a compensating increase in orbital angular momentum. This finite orbital angular momentum is transported by the boundary current.

We consider a minimal model of a two-dimensional chiral active fluid of dimers \cite{Kranthi_hydro_eqns, Cory_MD, my_PRL} composed of harmonically bonded monomers of diameter $\sigma$, in which all non-bonded monomers interact via the short-range, repulsive Weeks-Chandler-Andersen potential with characteristic energy $\epsilon$ \cite{WCA}. In our initial study, the dimers are confined by a frictionless circular wall of radius $R = 75 \, \sigma$ (Fig. \ref{fig:Fig1}). Each monomer is subjected to an active force of magnitude $F_a$ perpendicular to its harmonic bond, such that the dimers experience an active torque of average magnitude $F_a d_0$, where $d_0$ is the equilibrium bond length. The activity is quantified by a P\'{e}clet number ${\rm Pe} \equiv 2 \beta F_a d_0$, where $\beta$ is the inverse temperature times Boltzmann's constant (SI). 

We introduce in the SI a framework for deriving hydrodynamic balance laws in systems arbitrarily far from equilibrium inspired by Ref. \cite{Irving+Kirkwood} and modified for stochastic, rather than Hamiltonian, dynamics \cite{dean, underdamped_dean}. Using this framework, we obtain the hydrodynamic equations directly from our microscopic equations of motion. We find for an incompressible fluid of density $\rho_0$
\begin{equation}
\nabla \cdot {\boldsymbol u} = 0
\label{eqn:incompressible_continuity}
\end{equation}
\begin{equation}
\rho_0 D_t = - \frac{\gamma}{m} \rho_0 {\boldsymbol u} - \nabla p + \left( \eta_S + \eta_R \right) \nabla^2 {\boldsymbol u} + \eta_O \nabla^2 {\boldsymbol u}^*
\label{eqn:incompressible_linmom}
\end{equation}
and
\begin{equation}
\rho_0 D_t \Delta \mathcal{L} = - \frac{\gamma}{m} \rho_0 \Delta \mathcal{L} + 2 \eta_R \omega
\label{eqn:incompressible_spin}
\end{equation}
with ${\boldsymbol u}$ the velocity field, $\Delta \mathcal{L}$ the spin density relative to the constant bulk value in the unforced steady state, $D_t \equiv \partial_t + {\boldsymbol u} \cdot \nabla$ the material derivative, $\omega \equiv \left( \nabla^{3d} \times {\boldsymbol u}  \right) \cdot \hat{\boldsymbol z}$ the vorticity orthogonal to the plane of motion, $\gamma$ a linear substrate friction coefficient, and $m$ the monomer mass. These equations are statements of conservation of mass, linear momentum, and spin angular momentum, respectively. In addition to the ordinary shear viscosity $\eta_S$, two exotic transport coefficients appear in Eqs. \ref{eqn:incompressible_linmom} and \ref{eqn:incompressible_spin}: the rotational viscosity $\eta_R$, which couples the antisymmetric part of the stress tensor to the vorticity and has the effect of enhancing the diffusion of linear momentum, and the odd viscosity $\eta_O$ \cite{Avron, Kranthi_hydro_eqns, Banerjee, Kranthi1, bubble_dynamics_odd_viscosity, Vitellis_pretty_pictures, Liao_PDFmechanics, 3d_odd_viscosity}. The odd viscosity is linked to orthogonal fluid motion in Eq. \ref{eqn:incompressible_linmom} by the dual velocity ${\boldsymbol u}^* \equiv \varepsilon \cdot {\boldsymbol u}$, where $\varepsilon$ is the Levi-Civita tensor. Geometrically, a dual vector ${\boldsymbol a}^*$ is the vector ${\boldsymbol a}$ rotated clockwise by $90^\circ$. The microscopic model underpinning these hydrodynamic equations has been demonstrated to exhibit odd viscosity \cite{Cory_MD, my_PRL}. 

In confinement, the fluid spontaneously generates a boundary current (Fig. \ref{fig:Fig1}). The tangential velocity $v$ and spin anomaly profiles are derived by solving Eqs. \ref{eqn:incompressible_continuity}- \ref{eqn:incompressible_spin} subject to the integral statement of angular momentum conservation (SI)
\begin{equation}
\left\langle Q_{\rm hyd} \right\rangle_0 = \int_0^{R_{\rm hyd}} 2 \pi r {\rm d} r \, \rho_0 r v(r)
\label{eqn:integral_angmom}
\end{equation}
where $r$ is measured from the center of the confinement, $\langle \rangle_0$ indicates a nonequilibrium average in the unforced steady state, and $Q_{\rm hyd}$ is the total orbital angular momentum in the hydrodynamic region of the boundary current, which must be measured in simulation. The hydrodynamic description fails close to the interface where microscopic correlations persist (Figs. \ref{fig:Fig1}a and \ref{fig:Fig1}b) requiring us to define a hydrodynamic radius $R_{\rm hyd}$, which characterizes the region over which a hydrodynamic description is valid \cite{Bocquet_Barrat1994}. The hydrodynamic radius $R_{\rm hyd}$ is taken to coincide with the second density peak \cite{Chen_interface,my_JPCL}.

The solutions for the velocity and spin anomaly profiles $v(r)$ and $\Delta \mathcal{L}(r)$ subject to Eq. \ref{eqn:integral_angmom} are given by
\begin{equation}
v(r) = \frac{\kappa_{\gamma}}{2 \rho_0} \frac{\left\langle Q_{\rm hyd} \right\rangle_0}{\pi R_{\rm hyd}^2} e^{-\kappa_{\gamma} \left( R_{\rm hyd} - r \right)}
\label{eqn:v_BC}
\end{equation}
and
\begin{equation}
\Delta \mathcal{L} (r) = \frac{2 \eta_R}{\eta_S + \eta_R} \kappa_{\gamma}^{-1} v(r)
\label{eqn:DL_BC}
\end{equation}
with the inverse length scale $\kappa_{\gamma} \equiv \sqrt{(\gamma/m) \rho_0 / \left(\eta_S + \eta_R\right)}$. The theoretical profiles depend on only two transport coefficients, $\eta_S$ and $\eta_R$, allowing us to extract estimates of these parameters from fits of the velocity and spin profiles. We measure $\left\langle Q_{\rm hyd} \right\rangle_0 / \pi R_{\rm hyd}^2$ in each simulation, extract $\kappa_{\gamma}$ as the single free parameter in an exponential fit to the velocity profile, and estimate $\eta_R$ from a fit to $\Delta \mathcal{L}$ with $\kappa_{\gamma}^{-1}$ as the imposed decay length. This procedure collapses our data excellently (Figs. \ref{fig:Fig1}c and \ref{fig:Fig1}d), indicating the validity of our theory. This indicates that the flow is incompressible and well described by our first principles hydrodynamic equations.

\begin{figure}[t]
	\centerline{\includegraphics[width=8.5cm]{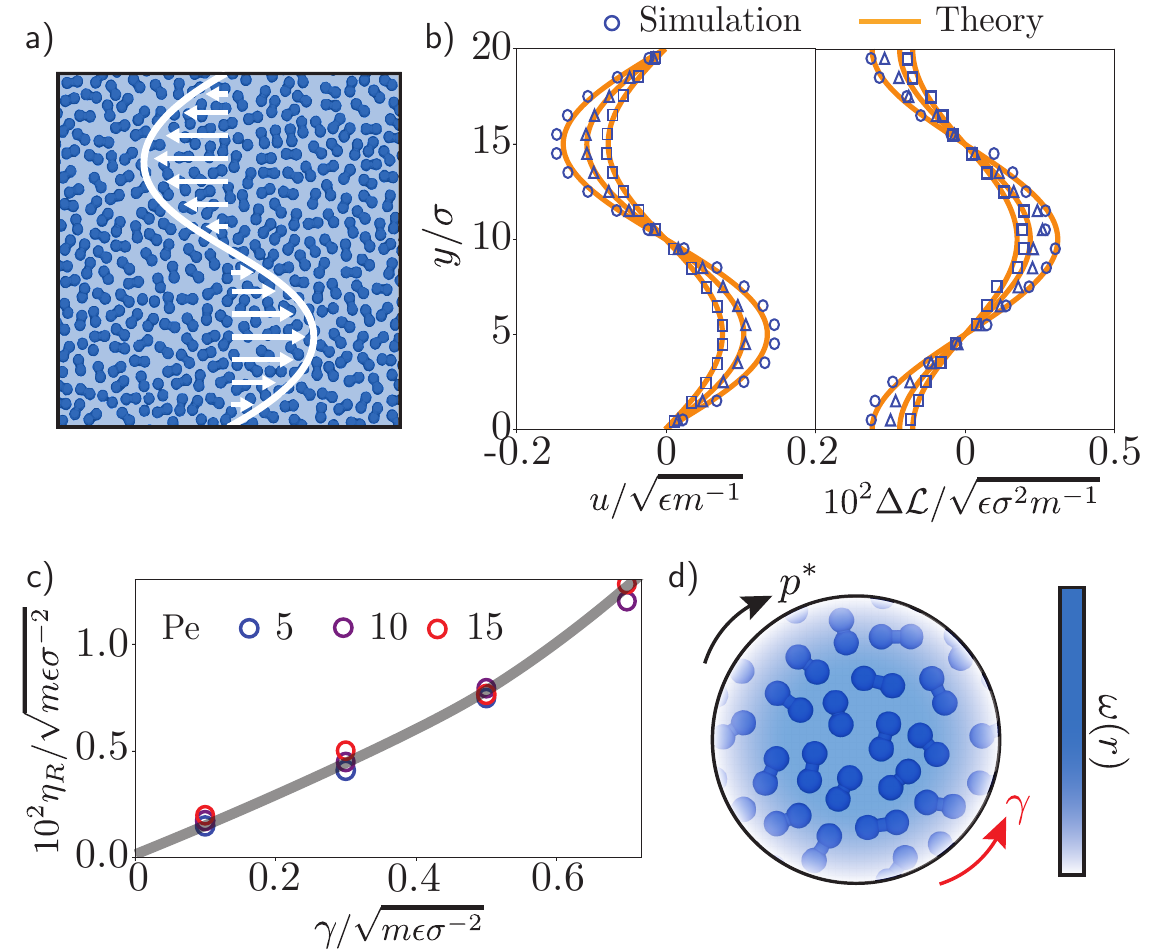}}
    \caption{Forced runs. a) Results are obtained for rotating dimers in a periodic box subjected to a sinusoidal applied force. b) Velocity and spin anomaly profile for $\gamma = 0.1, 0.3$ and $0.5 \, \sqrt{m \varepsilon \sigma^{-2}}$ and ${\rm Pe = 5}$. Theoretical comparisons are plotted with values of $\eta_S$ and $\eta_R$ inferred from boundary current. c) Estimated values of rotational viscosity from boundary current. d) Origin of rotational viscosity from local solid body rotation of fluid.}
\label{fig:Fig2}
\end{figure}

We test the values of $\eta_S$ and $\eta_R$ extracted from the boundary current simulations against forced runs conducted in a square periodic domain of side length $L = 100 \, \sigma$ in which the chiral active fluid is subjected to a force density ${\boldsymbol f}(y) = (\rho_0 F_0/m) {\rm sin} \left( k y \right)$, with $k = 10 \pi / L$. The anticipated theoretical profiles in this configuration depend on $\eta_S$ and $\eta_R$ and are derived in the SI. Figs. \ref{fig:Fig2}a and b show comparisons of forced simulation results to the theoretical profiles. The agreement is excellent over a range of microscopic parameters, further validating our hydrodynamic theory. This extends the previously observed consistency between linear response relations and forced runs \cite{Kranthi1, Cory_MD} observed for bulk-like homogeneous fluids in periodic simulations to the boundary value problem posed by the boundary current that forms spontaneously when the reference steady state is confined.

The rotational viscosity in our theory links the antisymmetric part of the stress tensor $\tau$ to the vorticity: $p^* \varepsilon \equiv \left( \tau - \tau^T \right)/2$, and $p^* = p^*_0 + \eta_R \omega$, with $p^*_0$ the bulk value of the antisymmetric stress in the unforced steady state (SI). This is distinct from how $\eta_R$ is typically defined \cite{etaR1, etaR1, etaR3, etaR4, etaR5, lubensky_etaR6}: $p^* = p^*_0 + \eta_R' \left( 2 \Omega - \omega \right)$ with $\Omega = 2m \mathcal{L} / I_0$ the dimer rotation frequency, and $I_0 \equiv md_0^2/2$ the dimer moment of inertia. This spin-vorticity coupling has previously been cited as the origin of boundary current formation \cite{lubensky_etaR6, waves_Kranthi, free_surface, crystal} and assumes a dissipation through friction induced by inter-rotor collisions that tends to result in a spin density $\mathcal{L} = (I_0/2m) (\omega/2)$. Because the repulsive interactions between monomers in our model are central, there is no such frictional dissipation, and this coupling need not be present \cite{etaR1,Kranthi1}.
We allow for the more general relation $p^* = p^*_0 + \eta_R \omega - \phi \Delta \mathcal{L}$, with $\phi$ a hypothetical transport coefficient linking the spin density to the antisymmetric stress. Our results show that $\phi$ does not play a significant role in the dynamics, and the coupling between spin and linear momentum is therefore one-way, with $\omega$ acting as a source for $\Delta \mathcal{L}$ in Eq. \ref{eqn:incompressible_spin}, but $\Delta \mathcal{L}$ playing no role in the linear momentum dynamics (Eq. \ref{eqn:incompressible_linmom}).

The coupling between stress and vorticity would be prohibited by the principle of objectivity in achiral fluids \cite{Kranthi1} but is permitted here. However, it is the breaking of Galilean, rather than parity, symmetry that results in a nonzero value of $\eta_R$. Galilean invariance is broken by a finite value of the substrate friction $\gamma$ and the corresponding introduction of a privileged substrate velocity. The substrate friction induced by the solid body rotation of a fluid element must be compensated by the net torque $p^*$ acting on the fluid element, leading to a finite $\eta_R$. This is corroborated in Figs. \ref{fig:Fig2}c and d, where we see that $\eta_R$ is only weakly dependent on $\rm Pe$, increases monotonically with $\gamma$, and extrapolates to zero as $\gamma$ vanishes.

Our hydrodynamic theory for the spontaneous boundary current reveals that this phenomenon is analogous to a well known phenomenon in coastal oceanography, the formation of a quasigeostrophic coastal current due to the local injection of freshwater into the coastal ocean (SI). The oceanographic coastal current forms to conserve freshwater \cite{Alex_ARFM, Alex_JFM}, analogous to the conservation of angular momentum that results in the formation of the boundary current (Eq. \ref{eqn:integral_angmom}). This analogy reveals that $\eta_O$ results in the spontaneous emergence of a Coriolis-like parameter $\eta_O \kappa_{\gamma}^2 / \rho_0$ proportional to the odd viscosity and the  square of the inverse confinement length $\kappa_{\gamma}^2$.

The direction of propgation of the coastal current is determined by that of coastally bound Kelvin waves, an oceanographic chiral edge mode \cite{continuum_topology, Kelvin_quantum_Hall} conjugate to the steady coastal curent. The chirality of this boundary mode is in accord with the band topographic notion of bulk-boundary correspondence: the presence of the topological order parameter $f$ breaks parity and time-reversal symmetry in the bulk of the fluid, introducing a band gap between bulk low frequency Rossby and high frequency Poincar\'{e} waves and therefore a nontrivial band topology \cite{science_topology}. This nontrivial topology is associated with an integer-valued topological invariant known as the Chern number, and the difference in these numbers across either the topological interface at the equator where $f$ changes sign or at continental boundaries where $f$  vanishes on the continent corresponds to the net number of permitted chiral modes at the boundary or interface \cite{science_topology, fuchs, boundary_waves, topology_soft_matter, Tauber_JFM}. The resultant edge bands exist in the bulk band gap \cite{fuchs}.

There is no imposed Coriolis force in our system, and the bulk band topology is therefore trivial \cite{boundary_waves} (SI). Bulk-boundary correspondence then predicts the absence of a net number of chiral edge modes. However, we are motivated by the spontaneous emergence of a Coriolis-like parameter in the boundary current solution to search for Kelvin-like chiral wave modes in our system. We search for one-dimensional, bound acoustic waves by relaxing the assumption of incompressibility in our hydrodynamic equations (SI) and allowing for linear perturbations $\Delta \rho$ and $\Delta v$ to the density and tangential velocity. This leads to the following system of equations:
\begin{equation}
\partial_t \Delta \rho = - \rho_0 \partial_y \Delta v
\label{eqn:unsteady_continuity}
\end{equation}
\begin{equation}
0 = - c^2 \partial_r^2 \Delta \rho + \eta_O \left( \partial_r^2 + \partial_y^2 \right) \Delta v
\label{eqn:unsteady_xmom}
\end{equation}
and
\begin{equation}
\rho_0 \partial_t \Delta v = - c^2 \partial_y \Delta \rho
\label{eqn:unsteady_ymom}
\end{equation}
with $c \equiv \sqrt{\partial_{\rho} p|_{\rho_0}}$ the bulk speed of sound when $\eta_O = 0$ and $y \equiv r \theta$ the circumferential coordinate. We have neglected frictional terms and coupling of the waves to the boundary current flow (SI). These equations are valid in the limit that the confinement radius $R$ becomes large compared to the wave thickness $\ell_r \equiv \kappa_r^{-1}$. This is the only regime in which boundary waves may exist in the chiral fluid (SI). As waves penetrate further into the bulk, their frequencies are shifted as a function of $r$ by the changing radius of curvature of the wave characteristic and an along-flow odd viscous pressure gradient that depends on the local radius of curvature. We demonstrate in the SI that no solution can exist to these equations when the frequency is radially dependent. This implies that the one-dimensional boundary waves devolve into two-dimensional bulk waves if their thicknesses become comparable to the confinement radius.

\begin{figure}[t]
	\centerline{\includegraphics[width=8.5cm]{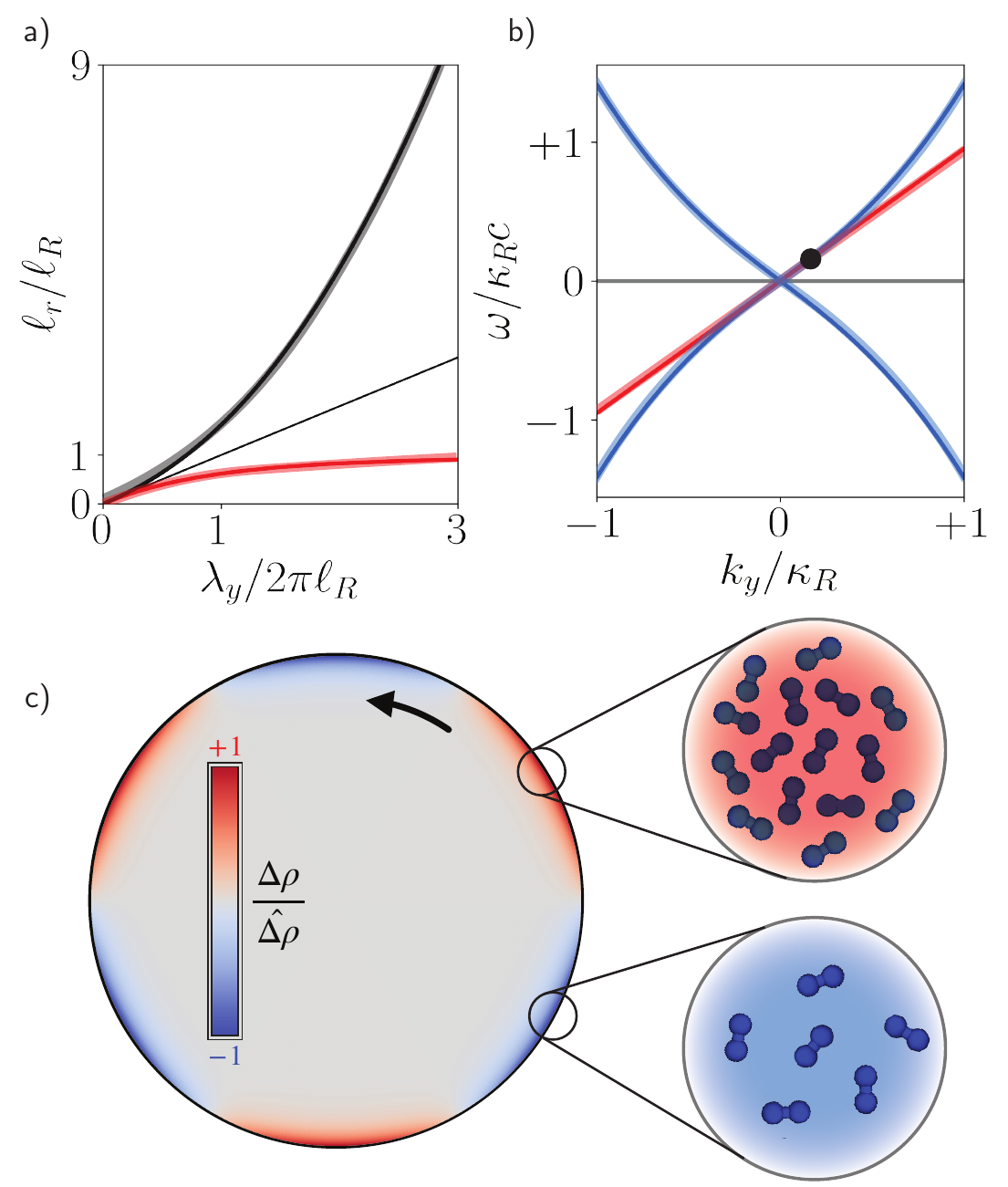}}
    \caption{Chiral Kelvin wave mode. a) Penetration length of Kelvin (red) and divergent (black) wave mode. b) Bulk dispersion relations (blue, gray) and Kelvin edge mode dispersion relation (red). 
     c) Theoretical structure of boundary-trapped acoustic Kelvin wave. Black dot in b indicate parameter values for wave plotted in c.}
\label{fig:Fig3}
\end{figure}

Eqs. \ref{eqn:unsteady_continuity} through \ref{eqn:unsteady_ymom} generically admit two solutions with frequencies $\omega_{\pm}^n = \pm S_{\eta_O} ck_y^n$, with $S_{\eta_O}$ the sign of the odd viscosity, and $k_y^n \equiv 2 \pi / \lambda_y = n/R$ the circumferential wave number. Here $n$ is a natural number, and $k_y^n = n/R$ is the periodicity condition in circular confinement. We plot in Fig. \ref{fig:Fig3}a the $e$-folding thicknesses of these modes as a function of circumferential wavelength $\lambda_y$ and normalized by the Rossby radius of deformation $\ell_R \equiv |\eta_O|/\rho_0 c$. The Rossby radius is defined generically as $\ell_R \equiv c/|f|$ in the oceanographic context and sets the thickness of the coastal Kelvin wave \cite{Kundu}. The confinement scale of the $-S_{\eta_O}$ mode diverges quadratically as $\lambda_y \rightarrow \infty$ and therefore must devolve into two dimensional bulk waves beyond some critical wavelength. This mode can only exist when $n \gg (1/2) \sqrt{1 + 2 \kappa_R R}$ (SI). That is, as the system size becomes large compared to the Rossby radius the negatively propagating mode is entirely eliminated, and we are left only with the Kelvin mode. The dispersion of this mode is compared to the bulk dispersion relations in Fig. \ref{fig:Fig3}b. Interestingly, this band fills the bulk band gap everywhere except at $k_y^n = 0$, where all of the frequencies vanish. 

Fig. \ref{fig:Fig3}a reveals that the thickness of the Kelvin mode saturates at the Rossby radius for large wavelengths, analogous to the coastal Kelvin wave. The Rossby radius $\ell_R \equiv |\eta_O|/\rho_0 c$  is proportional to our rotational parameter $\eta_O$, and thus increases in thickness with increasing rotational activity. The Kelvin acoustic mode is illustrated in Fig. \ref{fig:Fig3}c, which microscopically corresponds to alternating rarefaction or compaction of the active particles. Neither this mode nor the oceanographic coastal Kelvin wave are protected in the topological sense. In both cases, this is because their emergence is dependent on the no-flux boundary condition \cite{continuum_topology}. However, these waves are physically protected. As there is only one boundary wave mode with a sense of propagation dictated by the sign of the odd viscosity for system sizes $\kappa_R R \gg 1$, these waves must propagate without backscattering.

The Kelvin mode discovered here is qualitatively analogous to the chiral free surface mode observed in Ref. \cite{free_surface} for a system of rotating colloids, whose explanation was phenomenological. 
This suggests that the oceanographic analogy here is indeed useful in rationalizing the behavior of confined chiral active fluids. 
The mechanism that underpins this analogy is the spontaneous emergence of a Coriolis-like parameter in boundary trapped motions, and pushing this analogy further will reveal new physics analogous to, for example, two-dimensional instabilities and turbulence in the oceanographic context. This mode was revealed by a first principles derivation of the hydrodynamic balance equations, and this framework allows for the unambiguous derivation of hydrodynamic equations in systems far from equilibrium. Undoubtably, other exotic phenomena are waiting to be uncovered through analogous systematic evaluation of hydrodynamic equations of other classes of matter driven far from equilibrium.

\vspace{0.5cm}
\noindent{\bf Acknowledgments}
This work has been supported by NSF Grant No. CHE-1954580. A. R. P was also supported by the Heising-Simons Fellowship from the Kavli Energy Nanoscience Institute at UC Berkeley and D. T. L acknowledges support from the Alfred P. Sloan Foundation.
\vspace{0.5cm}

\end{document}


\title{Supplementary information for ``Emergent Kelvin waves in chiral active matter''}

\author{Anthony R. Poggioli}
\email{arpoggioli@berkeley.edu}
\affiliation{Department of Chemistry, University of California, Berkeley}
\affiliation{Kavli Energy NanoScience Institute, Berkeley, California}

\author{David T. Limmer}
\email{dlimmer@berkeley.edu}
\affiliation{Department of Chemistry, University of California, Berkeley}
\affiliation{Kavli Energy NanoScience Institute, Berkeley, California}
\affiliation{Materials Science Division, Lawrence Berkeley National Laboratory}
\affiliation{Chemical Science Division, Lawrence Berkeley National Laboratory}

\date{\today}

\maketitle

\tableofcontents

\section{I. Derivation of hydrodynamic equations}
\label{sec:hydro_derivation}

\subsection{Ia. Model description}

We consider $N_{\rm dimer}$ dimers confined to two-dimensions and composed of harmonically bonded monomers. In the forced runs, we consider $4100$ dimers in a periodic box of side length $L = 100 \, \sigma$, where $\sigma$ is the monomer diameter. In the boundary current simulations, $7068$ dimers are held in a circular confinement of radius $R = 75 \, \sigma$ using a short-range repulsive Weeks-Chandler-Andersen (WCA) wall \cite{WCA} with diameter and energy parameters $\sigma_{\rm wall} = \sigma$ and $\beta \epsilon_{\rm wall} = 10$, respectively. (See Eq. \ref{eqn:WCA} below for the form of the WCA potential.) Here, $\beta \equiv 1 / k_B T$ is the inverse of the Boltzmann constant times the temperature. Only the component of the wall-monomer interaction perpendicular to the wall is retained, meaning that the wall is frictionless and does not retard flow or dissipate angular momentum.

The stochastic equations of motion for the individual monomers, interpreted in the It\^{o} sense, are \cite{Cory_MD, my_PRL}
\begin{equation}
\dot{\boldsymbol x}_i^{\alpha} = \frac{{\boldsymbol p}_i^{\alpha}}{m}
\label{eqn:xeom_monomer}
\end{equation}
and
\begin{equation}
\dot{\boldsymbol p}_i^{\alpha} = - \frac{\gamma}{m} {\boldsymbol p}_i^{\alpha} + {\boldsymbol F}_{c,i}^{\alpha} + {\boldsymbol F}_{a,i}^{\alpha} + \sqrt{2 \gamma / \beta} {\boldsymbol \eta}_i^{\alpha} (t)
\label{eqn:peom_monomer}
\end{equation}
where $i \in \{1,...,N_{\rm dimer}\}$ indexes the dimers, $\alpha \in \{1,2\}$ indexes the monomers on dimer $i$, ${\boldsymbol x}_i^{\alpha}$ and ${\boldsymbol p}_i^{\alpha}$ are, respectively, the position and momentum of monomer $i$-$\alpha$, $m$ is the monomer mass, $\gamma$ is a linear (`dry') bath friction coefficient, ${\boldsymbol F}_{c,i}^{\alpha}$ is the sum of conservative forces acting on monomer $i$-$\alpha$, ${\boldsymbol F}_{a,i}^{\alpha}$ is the active force acting on the monomer, and ${\boldsymbol \eta}_i^{\alpha}(t)$ is a zero-mean, delta-correlated Guassian noise. The prefactor multiplying the noise $\sqrt{2\gamma/\beta}$ is determined by the fluctuation-dissipation theorem on the assumption of a local detailed balance between the underlying quiescent fluid bath and each system degree of freedom.

The linear friction assumed here is typically termed dry friction in the active matter literature \cite{joanny} and neglects any global flow developed within the underlying fluid bath \cite{Hosaka_mobility} as well as any inter-monomer hydrodynamic interactions potentially mediated by the fluid bath. To include such effects would greatly complicate the simulations and theory and likely result only in quantitative adjustments to our results. As such, we deem such considerations only appropriate if one wishes to quantitatively compare theoretical results directly to a particular experimental realization of an odd viscous fluid, and we therefore neglect such effects here, treating the underlying fluid bath as perfectly quiescent.

The total conservative force ${\boldsymbol F}_{c,i}^{\alpha}$ acting on a given monomer away from the confining wall is the sum of two terms: the harmonic bond between a given monomer and its intra-dimer pair and the repulsive WCA interaction between the given monomer and any other {\it non-bonded} monomers within the WCA cutoff, $r_{\rm WCA} = 2^{1/6} \sigma$. The harmonic and WCA potentials are given by, respectively,
\begin{equation}
u_{\rm harmonic} \left( d_i; k, d_0 \right) = k \left( d_i - d_0 \right)^2
\label{eqn:harmonic}
\end{equation}
and
\begin{equation}
\begin{aligned}
u_{\rm WCA} &\left( r_{ij}^{\alpha \beta}; \sigma, \epsilon \right) = \\
&\!\begin{cases}
4 \epsilon \left[ \left( \frac{\sigma}{r_{ij}^{\alpha \beta}} \right)^{12} - \left( \frac{\sigma}{r_{ij}^{\alpha \beta}} \right)^6 \right] + \epsilon, & r_{ij}^{\alpha \beta} < r_{\rm WCA} \\
0, & r_{ij}^{\alpha \beta} \geq r_{\rm WCA}
\end{cases}
\end{aligned}
\label{eqn:WCA}
\end{equation}
where $k$ is a spring stiffness constant, set equal to $100 \, \epsilon \, \sigma^{-2}$ in all of our simulations, ${\boldsymbol d}_i \equiv {\boldsymbol x}_{ii}^{12} \equiv {\boldsymbol x}_i^1 - {\boldsymbol x}_i^2$ is the instantaneous bond vector of length $d_i \equiv \left| {\boldsymbol d}_i \right|$, $d_0 = \sigma$ is the equilibrium bond length, set equal to the monomer diameter $\sigma$ in all of our simulations, $\beta \epsilon = 1$ with $\epsilon$ the WCA energy parameter, and $r_{ij}^{\alpha \beta} \equiv \left| {\boldsymbol x}_{ij}^{\alpha \beta} \right| \equiv \left| {\boldsymbol x}_i^{\alpha} - {\boldsymbol x}_j^{\beta} \right|$ is the distance between two non-bonded monomers $i$-$\alpha$ and $j$-$\beta$. We therefore have ${\boldsymbol F}_{c,i}^{\alpha} = - \nabla_{{\boldsymbol x}_i^{\alpha}} \left( u_{\rm harmonic} + u_{\rm WCA} \right)$.

Each monomer experiences an active force of fixed magnitude $F_a$ directed orthogonal to the instantaneous bond vector ${\boldsymbol d}_i$ associated with the dimer of which it is a member. This results in an instantaneous active torque of magnitude $F_a d_i(t)$ acting on dimer $i$ and an average active torque of magnitude $F_a d_0$, given that the density is not so large that it results in an average bond length $\left\langle d_i \right\rangle < d_0$. The direction of the active force is such that $F_a > 0$ corresponds to a counterclockwise active torque. That is, the active forces on monomers $1$ and $2$ of dimer $i$ are related to the bond vector by
\begin{equation}
{\boldsymbol F}_{a,i}^1 = - {\boldsymbol F}_{a,i}^2 = - F_a \hat{\boldsymbol d}_i^*
\label{eqn:Fai12}
\end{equation}
where $\hat{\boldsymbol d}_i^* \equiv \varepsilon \cdot \hat{\boldsymbol d}_i$ is the vector dual to $\hat{\boldsymbol d}_i$, $\hat{\boldsymbol d}_i \equiv {\boldsymbol d}_i / \left| {\boldsymbol d}_i \right|$ is the unit vector pointing along the instantaneous bond of dimer $i$, and $\varepsilon \equiv \hat{\boldsymbol x} \otimes \hat{\boldsymbol y} - \hat{\boldsymbol y} \otimes \hat{\boldsymbol x}$ is the antisymmetric symbol or Levi-Civita tensor in Cartesian coordinates. Action of $\boldsymbol \varepsilon$ on a vector results in clockwise rotation of that vector by an angle $\pi/2$.

The activity breaks parity and time-reversal symmetry in our system, leading to the exotic transport phenomena we investigate here and the breaking of detailed balance in the unforced steady state of our fluid. It is quantified by a P\'{e}clet number, defined as ${\rm Pe} \equiv 2 \beta F_a d_0$. All of our simulations are implemented in LAMMPS using Lennard-Jones units, in which the fundamental units of mass, time, and length are given by $m$, $\varepsilon$, and $\sigma$. All other units are derived from these quantities. The active force is implemented using code available at 
\cite{Cory_github}.

\subsection{Ib. Balance equations from microscopic model}

We now derive the equations for the balance of mass, linear momentum, and angular momentum in our system. We use what we call a stochastic Irving-Kirkwood approach, following the spirit of \cite{Irving+Kirkwood}, but modified by the consideration of stochastic, rather than Hamiltonian, dynamics \cite{dean, underdamped_dean}. In developing our hydrodynamic equations, we must consider not only conservation of mass and linear momentum, but of spin angular momentum as well. In order to do so, we must isolate the dimer center-of-mass dynamics from the spin dynamics. We do this by introducing the following decomposition of variables into center of mass and internal contributions:
\begin{equation}
{\boldsymbol x}_i \equiv \frac{{\boldsymbol x}_i^1 + {\boldsymbol x}_i^2}{2}; \quad {\boldsymbol d}_i \equiv {\boldsymbol x}_i^1 - {\boldsymbol x}_i^2
\label{eqn:x_decomp}
\end{equation}
\begin{equation}
{\boldsymbol p}_i \equiv {\boldsymbol p}_i^1 + {\boldsymbol p}_i^2; \quad {\boldsymbol \pi}_i \equiv {\boldsymbol p}_i^1 - {\boldsymbol p}_i^2
\label{eqn:p_decomp}
\end{equation}
and
\begin{equation}
{\boldsymbol F}_i \equiv {\boldsymbol F}_i^1 + {\boldsymbol F}_i^2; \quad {\boldsymbol \Delta F}_i \equiv {\boldsymbol F}_i^1 - {\boldsymbol F}_i^2
\label{eqn:F_decomp}
\end{equation}
We also define the dimer spin:
\begin{equation}
\ell_i \equiv \varepsilon : \frac{1}{2} {\boldsymbol d}_i \otimes {\boldsymbol \pi}_i
\label{eqn:elli_defn}
\end{equation}
where $A : B \equiv \sum_{ij} A_{ij} B_{ij}$ indicates a double contraction.

From Eqs. \ref{eqn:xeom_monomer} and \ref{eqn:peom_monomer}, we can derive the equations of motion for the dimer center of mass and spin:
\begin{equation}
\dot{\boldsymbol x}_i = \frac{{\boldsymbol p}_i}{2m}
\label{eqn:xeom_dimer}
\end{equation}
\begin{equation}
\dot{\boldsymbol p}_i = - \frac{\gamma}{m} {\boldsymbol p}_i + {\boldsymbol F}_{c,i} + \sqrt{4 \gamma / \beta} {\boldsymbol \eta}_i (t)
\label{eqn:peom_dimer}
\end{equation}
and
\begin{equation}
\dot{\ell}_i = - \frac{\gamma}{m} \ell_i + \varepsilon : \frac{1}{2} {\boldsymbol d}_i \otimes \Delta {\boldsymbol F}_{c,i} + F_a d_0 + \sqrt{\gamma d_0^2 / \beta} \mu_i(t)
\label{eqn:elli_eom}
\end{equation}
where ${\boldsymbol \eta}_i$ and $\mu_i$ are again zero-mean, delta-correlated Guassian noises. In deriving Eq. \ref{eqn:elli_eom} we have assumed that the dimer bonds may be approximated as being perfectly rigid with $\left\langle d_i \right\rangle = d_0$. We will make this assumption in all of the following development. Interestingly, the active forces exactly cancel in Eq. \ref{eqn:peom_dimer}; the activity is still implicitly present, however, through the non-central nature of ${\boldsymbol F}_{c,i}$ and the resultant nonzero average off-diagonal components of the virial ${\boldsymbol x}_i \otimes {\boldsymbol F}_i$. It is these terms that will lead to a nonvanishing antisymmetric stress and odd viscosity in our hydrodynamic equations.

We consider the fluctuating mass, linear momentum, and spin hydrodynamic densities, defined as
\begin{equation}
\rho \equiv \sum_i 2 m \delta \left( {\boldsymbol x} - {\boldsymbol x}_i \right)
\label{eqn:mass_density_defn}
\end{equation}
\begin{equation}
\rho {\boldsymbol u} \equiv \sum_i {\boldsymbol p}_i \delta \left( {\boldsymbol x} - {\boldsymbol x}_i \right)
\label{eqn:momentum_density_defn}
\end{equation}
and
\begin{equation}
\rho \mathcal{L} \equiv \sum_i \ell_i \delta \left( {\boldsymbol x} - {\boldsymbol x}_i \right)
\label{eqn:spin_density_defn}
\end{equation}
respectively. These all have the form
\begin{equation}
\rho \mathcal{A} \equiv \sum_i a_i \left( {\boldsymbol p}_i, \ell_i \right) \delta \left( {\boldsymbol x} - {\boldsymbol x}_i \right)
\label{eqn:density_general}
\end{equation}
with $a_i$ a single-particle observable linear in the momenta. Given this linearity, we have no diffusive terms proportional to the momentum Laplacians in the corresponding form of It\^{o}'s Lemma \cite{dean, underdamped_dean}, and we can take the local time-derivative of $\rho \mathcal{A}$ at a position ${\boldsymbol x}$ using the ordinary chain rule to obtain
\begin{equation}
\begin{split}
&\partial_t \left( \rho \mathcal{A} \right) + \nabla \cdot \frac{{\boldsymbol p}_i a_i}{2m} \delta \left( {\boldsymbol x} - {\boldsymbol x}_i \right) \\
&\quad = \sum_i \left[ \dot{\boldsymbol p}_i \cdot \partial_{{\boldsymbol p}_i} a_i \delta \left( {\boldsymbol x} - {\boldsymbol x}_i \right) + \dot{\ell}_i \partial_{\ell_i} a_i \delta \left( {\boldsymbol x} - {\boldsymbol x}_i \right) \right]
\end{split}
\label{eqn:general_hydro1}
\end{equation}
where $\nabla$ denotes the gradient in the coordinate ${\boldsymbol x}$. This result is very nearly our general form of the fluctuating balance equation, except that we must first rewrite the second term on the LHS in the traditional advective form. This is done by the following manipulation of the product $\rho {\boldsymbol u} \rho \mathcal{A}$ \cite{dean}:
\begin{equation}
\begin{split}
\rho {\boldsymbol u} \rho \mathcal{A} &= \sum_{ij} {\boldsymbol p}_i a_j \delta \left( {\boldsymbol x} - {\boldsymbol x}_i \right) \delta \left( {\boldsymbol x} - {\boldsymbol x}_j \right) \\
&= \sum_j \delta \left( {\boldsymbol x} - {\boldsymbol x}_j \right) \sum_i {\boldsymbol p}_i a_i \delta \left( {\boldsymbol x} - {\boldsymbol x}_i \right) \\
&= \rho \sum_i \frac{{\boldsymbol p}_i a_i}{2m} \delta \left( {\boldsymbol x} - {\boldsymbol x}_i \right)
\end{split}
\label{eqn:density_product}
\end{equation}
where in the second equality we have noted that $\delta \left( {\boldsymbol x} - {\boldsymbol x}_i \right) \delta \left( {\boldsymbol x} - {\boldsymbol x}_j \right) = \delta \left( {\boldsymbol x} - {\boldsymbol x}_i \right) \delta \left( {\boldsymbol x}_i - {\boldsymbol x}_j \right)$ and assumed that $\delta \left( {\boldsymbol x}_i - {\boldsymbol x}_j \right) \propto \delta_{ij}$ -- {\it i.e.}, the centers of mass of two distinct dimers cannot overlap due to interparticle repulsion \cite{dean, Lubensky_PB}. Inserting this result into Eq. \ref{eqn:general_hydro1} gives
\begin{equation}
\begin{split}
&\partial_t \left( \rho \mathcal{A} \right) + \nabla \cdot \left( {\boldsymbol u} \rho \mathcal{A} \right) \\
&\quad = \sum_i \left[ \dot{\boldsymbol p}_i \cdot \partial_{{\boldsymbol p}_i} a_i \delta \left( {\boldsymbol x} - {\boldsymbol x}_i \right) + \dot{\ell}_i \partial_{\ell_i} a_i \delta \left( {\boldsymbol x} - {\boldsymbol x}_i \right) \right]
\end{split}
\label{eqn:general_hydro_final}
\end{equation}
This is one of the two fundamental results that we will use in deriving the hydrodynamic balance equations.

The other will be used to rewrite interaction terms in the hydrodynamic equations in conservative form and comes simply from a Taylor expansion of the difference of delta functions
\begin{equation}
\delta \left( {\boldsymbol x} - {\boldsymbol x}_i \right) - \delta \left( {\boldsymbol x} - {\boldsymbol x}_j \right) = - \nabla \cdot \left[ {\boldsymbol x}_{ij} b_{ij} \left( {\boldsymbol x} - {\boldsymbol x}_i; {\boldsymbol x}_{ij} \right) \right]
\label{eqn:delta_diff}
\end{equation}
where ${\boldsymbol x}_{ij} \equiv {\boldsymbol x}_i - {\boldsymbol x}_j$, and the function $b_{ij}$ is defined as
\begin{equation}
\begin{split}
b_{ij} \left( {\boldsymbol x} - {\boldsymbol x}_i; {\boldsymbol x}_{ij} \right) &\equiv \sum_{n = 1}^{\infty} \frac{\left\langle {\boldsymbol x}_{ij}^{\otimes n -1}, \nabla^{\otimes n -1} \right\rangle}{n!} \delta \left( {\boldsymbol x} - {\boldsymbol x}_i \right) \\
&\xrightarrow[\sigma/L \rightarrow \infty]{} \delta \left( {\boldsymbol x} - {\boldsymbol x}_i \right)
\end{split}
\label{eqn:bond_function}
\end{equation}
with $\langle A , B \rangle \equiv \sum_{ij \cdot \cdot \cdot k} A_{ij \cdot \cdot \cdot k} B_{ij \cdot \cdot \cdot k}$ a tensor inner product. We have noted that, in the limit that the hydrodynamic length scales of interest $L$ are large compared to the microscopic length scales of interaction/correlation $\sigma$, the bond function reduces simply to a delta function. This assumption of a separation of scales is the {\it hydrodynamic limit}. It allows us to meaningfully partition interaction densities into body terms and surface fluxes, and we will assume that it holds in everything that follows.

We apply Eq. \ref{eqn:general_hydro_final} to the mass and linear momentum densities defined in Eqs. \ref{eqn:mass_density_defn} and \ref{eqn:momentum_density_defn}, respectively, to find
\begin{equation}
\partial_t \rho + \nabla \cdot \left( {\boldsymbol u} \rho \right) = 0
\label{eqn:continuity_initial}
\end{equation}
and
\begin{equation}
\begin{split}
&\partial_t \left( \rho {\boldsymbol u} \right) + \nabla \cdot \left( {\boldsymbol u} \otimes \rho {\boldsymbol u} \right) = \sum_i \dot{\boldsymbol p}_i \delta \left( {\boldsymbol x} - {\boldsymbol x}_i \right) \\
&= - \frac{\gamma}{m} \rho {\boldsymbol u} + \sum_i {\boldsymbol F}_{c,i} \delta \left( {\boldsymbol x} - {\boldsymbol x}_i \right) + \sqrt{4 \gamma / \beta} \sum_i {\boldsymbol \eta}_i \delta \left( {\boldsymbol x} - {\boldsymbol x}_i \right)
\end{split}
\label{eqn:momentum_initial}
\end{equation}
where in the second equality we have inserted the mesoscopic equation of motion for the dimer center of mass, Eq. \ref{eqn:peom_dimer}. Eq. \ref{eqn:momentum_initial} will only be of a consistent hydrodynamic form if we can write the density of conservative forces as a conservative (divergence) term and close the sum of microscopic noises in the hydrodynamic variables. We begin with the latter. The noise term is a sum of Gaussians and accordingly only its first and second cumulants will be nonvanishing. Given that we are working in the It\^{o} convention, they are calculated as \cite{dean}
\[
\begin{split}
&\left\langle \sqrt{4 \gamma / \beta} \sum_i {\boldsymbol \eta}_i(t) \delta \left( {\boldsymbol x} - {\boldsymbol x}_i(t) \right) \right\rangle \\
&\quad \quad \quad \quad \quad \quad = \sqrt{4 \gamma / \beta} \left\langle {\boldsymbol \eta}_i (t) \right\rangle \left\langle \delta \left( {\boldsymbol x} - {\boldsymbol x}_i(t) \right) \right\rangle = 0
\end{split}
\label{eqn:first_cumulant}
\]
and
\[
\begin{split}
&\left\langle \frac{4 \gamma}{\beta} \sum_i {\boldsymbol \eta}_i(t) \delta \left( {\boldsymbol x} - {\boldsymbol x}_i (t) \right) \otimes \sum_j {\boldsymbol \eta}_j (t') \delta \left( {\boldsymbol x}' - {\boldsymbol x}_j (t') \right) \right\rangle \\
&= \frac{4 \gamma}{\beta} \sum_{ij} \left\langle {\boldsymbol \eta}_i(t) \otimes {\boldsymbol \eta}_j(t') \right\rangle \left\langle \delta \left( {\boldsymbol x} - {\boldsymbol x}_i(t) \right) \delta \left( {\boldsymbol x}' - {\boldsymbol x}_j(t') \right) \right\rangle \\
&= \frac{4 \gamma}{\beta} \sum_{ij} \mathbbm{1} \delta_{ij} \delta \left( t - t' \right) \left\langle \delta \left( {\boldsymbol x} - {\boldsymbol x}_i(t) \right) \delta \left( {\boldsymbol x}' - {\boldsymbol x}_i(t) \right) \right\rangle \\
&= \frac{4 \gamma}{\beta} \mathbbm{1} \delta \left( t - t' \right) \delta \left( {\boldsymbol x} - {\boldsymbol x}' \right) \left\langle \sum_i \delta \left( {\boldsymbol x} - {\boldsymbol x}_i \right) \right\rangle \\
&= \frac{2 \gamma \left\langle \rho \right\rangle}{m \beta} \mathbbm{1} \delta \left( t - t' \right) \delta \left( {\boldsymbol x} - {\boldsymbol x}' \right)
\end{split}
\label{eqn:second_cumulant}
\]
where angled brackets indicate a noise average, and we have made use of the correlator $\left\langle {\boldsymbol \eta}_i (t) \otimes {\boldsymbol \eta}_j (t') \right\rangle = \mathbbm{1} \delta_{ij} \delta \left( t - t' \right)$ and the noise-averaged form of Eq. \ref{eqn:mass_density_defn}. The noise term appearing in Eq. \ref{eqn:momentum_initial} may therefore be written as
\begin{equation}
\sum_i \sqrt{4 \gamma / \beta} {\boldsymbol \eta}_i(t) \delta \left( {\boldsymbol x} - {\boldsymbol x}_i(t) \right) = \sqrt{ \frac{2 \gamma \rho}{m \beta} } {\boldsymbol \xi} \left( {\boldsymbol x}, t \right)
\label{eqn:noise_momentum_eqn}
\end{equation}
where ${\boldsymbol \xi} \left( {\boldsymbol x}, t \right)$ is a delta-correlated space-time Gaussian noise. We see that we have now closed the noise in the hydrodynamic variables at the expense of introducing a noise multiplicative in the density \cite{dean}.

Finally, we may rewrite the interaction force density $\sum_i {\boldsymbol F}_{c,i} \delta \left( {\boldsymbol x} - {\boldsymbol x}_i \right)$ appearing in Eq. \ref{eqn:momentum_initial} using Eqs. \ref{eqn:delta_diff} and \ref{eqn:bond_function} in the hydrodynamic limit as \cite{Irving+Kirkwood}
\[
\begin{split}
&\sum_i {\boldsymbol F}_{c,i} \delta \left( {\boldsymbol x} - {\boldsymbol x}_i \right) = \sum_{ij} {\boldsymbol F}_{c,ij} \delta \left( {\boldsymbol x} - {\boldsymbol x}_i \right) \\
&= \frac{1}{2} \sum_{ij} {\boldsymbol F}_{c,ij} \delta \left( {\boldsymbol x} - {\boldsymbol x}_i \right) - \frac{1}{2} \sum_{ij} {\boldsymbol F}_{c,ji} \delta \left( {\boldsymbol x} - {\boldsymbol x}_i \right) \\
&= \frac{1}{2} \sum_{ij} {\boldsymbol F}_{c,ij} \left[ \delta \left( {\boldsymbol x} - {\boldsymbol x}_i \right) - \delta \left( {\boldsymbol x} - {\boldsymbol x}_j \right) \right] \\
&\xrightarrow[\sigma/L \rightarrow 0]{} \nabla \cdot \left[ - \frac{1}{2} \sum_{ij} {\boldsymbol x}_{ij} \otimes {\boldsymbol F}_{c,ij} \delta \left( {\boldsymbol x} - {\boldsymbol x}_i \right) \right]
\end{split}
\]
where in the second equality we have applied Newton's Third Law and in the third we have simply relabeled dummy indices. Motivated by this result, we define the Irving-Kirkwood virial stress as
\begin{equation}
\tau^V \equiv - \frac{1}{2} \sum_{ij} {\boldsymbol x}_{ij} \otimes {\boldsymbol F}_{c,ij} \delta \left( {\boldsymbol x} - {\boldsymbol x}_i \right)
\label{eqn:virial_defn}
\end{equation}

Inserting Eq. \ref{eqn:noise_momentum_eqn} and the divergence of Eq. \ref{eqn:virial_defn} into Eq. \ref{eqn:momentum_initial} gives
\begin{equation}
\partial_t \left( \rho {\boldsymbol u} \right) + \nabla \cdot \left( {\boldsymbol u} \otimes \rho {\boldsymbol u} \right) = - \frac{\gamma}{m} \rho {\boldsymbol u} + \nabla \cdot \tau^V + \sqrt{\frac{2 \gamma \rho}{m \beta}} {\boldsymbol \xi} \left( {\boldsymbol x}, t \right)
\label{eqn:momentum_fluctuating}
\end{equation}
our fluctuating balance equation for the linear momentum.

We follow a similar procedure to obtain the spin balance equation from Eqs. \ref{eqn:elli_eom} and \ref{eqn:general_hydro_final}. We find
\begin{equation}
\begin{split}
&\partial_t \left( \rho \mathcal{L} \right) + \nabla \cdot \left( {\boldsymbol u} \rho \mathcal{L} \right) \\
&\quad \quad= - \frac{\gamma}{m} \rho \mathcal{L} + \frac{F_a d_0}{2m} \rho + \sum_i \varepsilon : \frac{1}{2} {\boldsymbol d}_i \otimes \Delta {\boldsymbol F}_{c,i} \delta \left( {\boldsymbol x} - {\boldsymbol x}_i \right) \\
&\quad \quad \quad \quad \quad \quad \quad \quad + \sum_i \sqrt{\gamma d_0^2 / \beta} \mu_i(t) \delta \left( {\boldsymbol x} - {\boldsymbol x}_i \right)
\end{split}
\label{eqn:spin_initial}
\end{equation}
The noise term is trivially rewritten using the procedure outlined above as
\begin{equation}
\sum_i \sqrt{\gamma d_0^2 / \beta} \mu_i \delta \left( {\boldsymbol x} - {\boldsymbol x}_i \right) = \sqrt{\frac{\gamma d_0^2 \rho}{2 m \beta}} \zeta \left( {\boldsymbol x}, t \right)
\label{eqn:noise_spin}
\end{equation}
with $\zeta \left( {\boldsymbol x}, t \right)$ again a delta-correlated space-time Gaussian noise. As before, we have closed the noise in the hydrodynamic variables at the expense of making it multiplicative in the density.

The final step in deriving the fluctuating balance law for the spin density is rewriting the interaction density term appearing in Eq. \ref{eqn:spin_initial} in an explicitly conservative form using Eqs. \ref{eqn:delta_diff} and \ref{eqn:bond_function} in the hydrodynamic limit. This is a fair bit more involved than the corresponding calculation for the linear momentum equation. We begin by noting that the interaction term may be generically decomposed in the hydrodynamic limit as the sum of a line flux $\nabla \cdot {\boldsymbol c}^V$ and a surface source density $s^V$:
\begin{equation}
\sum_i \varepsilon : \frac{1}{2} {\boldsymbol d}_i \otimes \Delta {\boldsymbol F}_{c,i} \delta \left( {\boldsymbol x} - {\boldsymbol x}_i \right) = \nabla \cdot {\boldsymbol c}^V + s^V
\label{eqn:spin_interaction_generic_decomp}
\end{equation}
We also note that the {\it total} internal torque density may be written as
\begin{equation}
\begin{split}
\sum_{i \alpha} \varepsilon : {\boldsymbol x}_i^{\alpha} \otimes &{\boldsymbol F}_{c,i}^{\alpha} \delta \left( {\boldsymbol x} - {\boldsymbol x}_i \right) = \sum_i \varepsilon : {\boldsymbol x}_i \otimes {\boldsymbol F}_{c,i} \delta \left( {\boldsymbol x} - {\boldsymbol x}_i \right) \\
&+ \sum_i \varepsilon : \frac{1}{2} {\boldsymbol d}_i \otimes \Delta {\boldsymbol F}_{c,i} \delta \left( {\boldsymbol x} - {\boldsymbol x}_i \right)
\end{split}
\label{eqn:total_torque_density_decomp}
\end{equation}
Because all of the individual conservative interactions in the system are central the integral of the total internal torque density over the entire system must vanish. Likewise, the contribution to the integral over any `macroscopic' subvolume within the system from interactions interior to that volume will tend to cancel on average leaving only a contribution scaling with the size of the subvolume boundary. Hence, the total internal torque density must be given by a term of the form $\nabla \cdot {\boldsymbol d}$ for some vector field ${\boldsymbol d}$ -- the exact form of which we will give below. From this condition and Eqs. \ref{eqn:spin_interaction_generic_decomp} and \ref{eqn:total_torque_density_decomp} we therefore have
\begin{equation}
\nabla \cdot {\boldsymbol c}^V + s^V + \sum_i \varepsilon : {\boldsymbol x}_i \otimes {\boldsymbol F}_{c,i} \delta \left( {\boldsymbol x} - {\boldsymbol x}_i \right) = \nabla \cdot {\boldsymbol d}
\label{eqn:spin_decomp_2}
\end{equation}
In light of the delta function appearing in the third term on the LHS, we may rewrite this term as
\begin{equation}
\begin{split}
\sum_i \varepsilon : {\boldsymbol x}_i \otimes {\boldsymbol F}_{c,i} \delta \left( {\boldsymbol x} - {\boldsymbol x}_i \right) &= \varepsilon : {\boldsymbol x} \otimes \sum_i {\boldsymbol F}_{c,i} \delta \left( {\boldsymbol x} - {\boldsymbol x}_i \right) \\
&= \varepsilon : {\boldsymbol x} \otimes \nabla \cdot \tau^V
\end{split}
\label{eqn:tauV_cross_term}
\end{equation}
where in the second equality we have recognized the interaction force density as the divergence of the virial stress. (See Eq. \ref{eqn:virial_defn} and the preceding calculation and discussion.) By direct calculation in Cartesian coordinates, one may show that $\varepsilon : {\boldsymbol x} \otimes \nabla \cdot \tau^V = - \varepsilon : \tau^V - \nabla \cdot \left[ \left( \tau^V \otimes {\boldsymbol x} \right) : \varepsilon \right]$. We therefore have from Eqs. \ref{eqn:spin_decomp_2} and \ref{eqn:tauV_cross_term}
\begin{equation}
\nabla \cdot \left[ {\boldsymbol c}^V - \left( \tau^V \otimes {\boldsymbol x} \right) : \varepsilon \right] + \left( s^V - \varepsilon : \tau_V \right) = \nabla \cdot {\boldsymbol d}
\label{eqn:term_balance_couple_stress}
\end{equation}
from which we deduce
\begin{equation}
s^V = \varepsilon : \tau^V \equiv 2 p^*
\label{eqn:source_term_spin_eqn}
\end{equation}
where $p^* \equiv \left( \tau^V_{xy} - \tau^V_{yx} \right)/2$ is the antisymmetric component of the stress tensor, and
\begin{equation}
\begin{split}
\nabla \cdot {\boldsymbol c}^V &= \nabla \cdot \left[ {\boldsymbol d} + \left( \tau^V \otimes {\boldsymbol x} \right) : \varepsilon \right] \\
&= \sum_{i\alpha} \varepsilon : {\boldsymbol x}_i^{\alpha} \otimes {\boldsymbol F}_i^{\alpha} \delta \left( {\boldsymbol x} - {\boldsymbol x}_i \right) + \nabla \cdot \left[ \left( \tau^V \otimes {\boldsymbol x} \right) : \varepsilon \right]
\end{split}
\label{eqn:couple_stress_divergence}
\end{equation}
where we have recalled that $\nabla \cdot {\boldsymbol d}$ is by definition the total internal torque density.

The total torque density may be explicitly rewritten in divergent form using Newton's Third Law, the central nature of the pairwise interactions ${\boldsymbol F}_{ij}^{\alpha \beta}$, and Eqs. \ref{eqn:delta_diff} and \ref{eqn:bond_function} in the hydrodynamic limit. The result of this calculation leads us to introduce the following definition for ${\boldsymbol d}$:
\begin{equation}
{\boldsymbol d} \equiv - \frac{1}{2} \sum_{ij\alpha\beta} {\boldsymbol x}_{ij} \left( \varepsilon : {\boldsymbol x}_i^{\alpha} \otimes {\boldsymbol F}_{ij}^{\alpha\beta} \right) \delta \left( {\boldsymbol x} - {\boldsymbol x}_i \right)
\label{eqn:d_definition}
\end{equation}
The term $\left( \tau^V \otimes {\boldsymbol x} \right) : \varepsilon$ appearing in Eq. \ref{eqn:couple_stress_divergence} may be rewritten using Eq. \ref{eqn:virial_defn} as
\begin{equation}
\begin{split}
\left( \tau^V \otimes {\boldsymbol x} \right) : \varepsilon &= \frac{1}{2} \sum_{ij\alpha\beta} {\boldsymbol x}_i \left( \varepsilon : {\boldsymbol x} \otimes {\boldsymbol F}_{ij}^{\alpha\beta} \right) \delta \left( {\boldsymbol x} - {\boldsymbol x}_i \right) \\
&= \frac{1}{2} \sum_{ij\alpha\beta} {\boldsymbol x}_i \left( \varepsilon : {\boldsymbol x}_i \otimes {\boldsymbol F}_{ij}^{\alpha\beta} \right) \delta \left( {\boldsymbol x} - {\boldsymbol x}_i \right)
\end{split}
\end{equation}
from which we finally obtain for ${\boldsymbol c}^V$
\begin{equation}
{\boldsymbol c}^V \equiv - \frac{1}{2} \sum_{ij\alpha\beta} {\boldsymbol x}_{ij} \left[ \varepsilon : \left( {\boldsymbol x}_i^{\alpha} - {\boldsymbol x}_i \right) \otimes {\boldsymbol F}_{ij}^{\alpha\beta} \right] \delta \left( {\boldsymbol x} - {\boldsymbol x}_i \right)
\label{eqn:couple_stress_defn}
\end{equation}
Thus, inserting \ref{eqn:noise_spin}, \ref{eqn:spin_interaction_generic_decomp}, and \ref{eqn:source_term_spin_eqn} into Eq. \ref{eqn:spin_initial}, we finally obtain for the fluctuating spin density balance equation
\begin{equation}
\begin{split}
&\partial_t \left( \rho \mathcal{L} \right) + \nabla \cdot \left( {\boldsymbol u} \rho \mathcal{L} \right) \\
&= - \frac{\gamma}{m} \rho \mathcal{L} + \frac{F_a d_0}{2m} \rho + 2 p^* + \nabla \cdot {\boldsymbol c}^V + \sqrt{\frac{\gamma d_0^2 \rho}{2 m \beta}} \zeta \left( {\boldsymbol x}, t \right)
\end{split}
\label{eqn:spin_fluctuating}
\end{equation}

\subsection{Ic. Noise-averaged balance equations}

We derive in this section the noise-averaged forms of the balance equations, Eqs. \ref{eqn:continuity_initial}, \ref{eqn:momentum_fluctuating}, and \ref{eqn:spin_fluctuating}. We will indicate fluctuating quantities with primes and omit angled brackets on the corresponding noise-averaged hydrodynamic variables. We start by defining our noise-averaged hydrodynamic densities in the usual way as \cite{Irving+Kirkwood}
\begin{equation}
\rho \equiv \left\langle \sum_i 2 m \delta \left( {\boldsymbol x} - {\boldsymbol x}_i \right) \right\rangle
\label{eqn:hydro_mass_defn}
\end{equation}
\begin{equation}
\rho {\boldsymbol u} \equiv \left\langle \sum_i {\boldsymbol p}_i \delta \left( {\boldsymbol x} - {\boldsymbol x}_i \right) \right\rangle
\label{eqn:hydro_velocity_defn}
\end{equation}
and
\begin{equation}
\rho \mathcal{L} \equiv \left\langle \sum_i \ell_i \delta \left( {\boldsymbol x} - {\boldsymbol x}_i \right) \right\rangle
\label{eqn:hydro_spin_defn}
\end{equation}
Then, comparing Eqs. \ref{eqn:mass_density_defn} through \ref{eqn:spin_density_defn} to Eqs. \ref{eqn:hydro_mass_defn} through \ref{eqn:hydro_spin_defn}, we have by definition $\left\langle \rho' \right\rangle \equiv \rho$, $\left\langle \rho' {\boldsymbol u}' \right\rangle \equiv \rho {\boldsymbol u}$, and $\left\langle \rho' \mathcal{L}' \right\rangle \equiv \rho \mathcal{L}$.

Averaging Eqs. \ref{eqn:continuity_initial}, \ref{eqn:momentum_fluctuating}, and \ref{eqn:spin_fluctuating} over the noise gives
\begin{equation}
\partial_t \rho + \nabla \cdot \left( \rho {\boldsymbol u} \right) = 0
\label{eqn:continuity_noise_averaged}
\end{equation}
\begin{equation}
\partial_t \left( \rho {\boldsymbol u} \right) + \nabla \cdot \left\langle {\boldsymbol u}' \otimes \rho' {\boldsymbol u}' \right\rangle = - \frac{\gamma}{m} \rho {\boldsymbol u} + \nabla \cdot \tau^V
\label{eqn:momentum_noise_averaged1}
\end{equation}
and
\begin{equation}
\partial_t \left( \rho \mathcal{L} \right) + \nabla \cdot \left\langle {\boldsymbol u}' \rho' \mathcal{L}' \right\rangle = - \frac{\gamma}{m} \rho \mathcal{L} + \frac{F_a d_0}{2m} \rho + 2 p^* + \nabla \cdot {\boldsymbol c}^V
\label{eqn:spin_noise_averaged1}
\end{equation}

We rewrite the average of the nonlinear advective momentum flux appearing in Eq. \ref{eqn:momentum_noise_averaged1} by first defining the kinetic stress tensor
\begin{equation}
\tau^K \equiv - \left\langle \sum_i \frac{\left( {\boldsymbol p}_i - 2 m {\boldsymbol u} \right) \otimes \left( {\boldsymbol p}_i - 2 m {\boldsymbol u} \right) }{2m} \delta \left( {\boldsymbol x} - {\boldsymbol x}_i \right) \right\rangle
\label{eqn:tauk_defn}
\end{equation}
Then expanding terms gives
\[
\begin{split}
\tau^K &= - \left\langle \sum_i \frac{{\boldsymbol p}_i \otimes {\boldsymbol p}_i}{2 m} \delta \left( {\boldsymbol x} - {\boldsymbol x}_i \right) \right\rangle + {\boldsymbol u} \otimes \rho {\boldsymbol u} \\
&= - \left\langle {\boldsymbol u}' \otimes \rho' {\boldsymbol u}' \right\rangle + {\boldsymbol u} \otimes \rho {\boldsymbol u}
\end{split}
\]
where the second equality follows from Eq. \ref{eqn:density_product}. Inserting this result into Eq. \ref{eqn:momentum_noise_averaged1} gives
\begin{equation}
\partial_t \left( \rho {\boldsymbol u} \right) + \nabla \cdot \left( {\boldsymbol u} \otimes \rho {\boldsymbol u} \right) = - \frac{\gamma}{m} \rho {\boldsymbol u} + \nabla \cdot \tau
\label{eqn:momentum_noise_averaged2}
\end{equation}
with $\tau \equiv \tau^V + \tau^K$.

We likewise introduce the kinetic couple stress
\begin{equation}
{\boldsymbol c}^K \equiv - \sum_i \frac{\left( \ell_i - 2 m \mathcal{L} \right) \left( {\boldsymbol p}_i - 2 m {\boldsymbol u} \right)}{2m} \delta \left( {\boldsymbol x} - {\boldsymbol x}_i \right)
\label{eqn:cK_defn}
\end{equation}
expand this definition to find
\[
{\boldsymbol c}^K = - \left\langle {\boldsymbol u}' \rho' \mathcal{L}' \right\rangle + {\boldsymbol u} \rho \mathcal{L}
\]
and insert this result into Eq. \ref{eqn:spin_noise_averaged1} to obtain
\begin{equation}
\partial_t \left( \rho \mathcal{L} \right) + \nabla \cdot \left( {\boldsymbol u} \rho \mathcal{L} \right) = - \frac{\gamma}{m} \rho \mathcal{L} + \frac{F_a d_0}{2 m} \rho + 2 p^* + \nabla \cdot {\boldsymbol c}
\label{eqn:spin_noise_averaged2}
\end{equation}
with ${\boldsymbol c} \equiv {\boldsymbol c}^K + {\boldsymbol c}^V$.

We rewrite Eqs. \ref{eqn:continuity_noise_averaged}, \ref{eqn:momentum_noise_averaged2}, and \ref{eqn:spin_noise_averaged2} by introducing the material derivative operator
\begin{equation}
D_t \equiv \partial_t + {\boldsymbol u} \cdot \nabla
\label{eqn:material_derivative}
\end{equation}
which returns the total time derivative following a fluid element. Then, using Eq. \ref{eqn:continuity_noise_averaged}, we can rewrite the sum of the local unsteady and advective flux divergence terms for a generic density $\mathcal{A}$ as
\begin{equation}
\begin{split}
\partial_t \left( \rho \mathcal{A} \right) + \nabla \cdot \left( {\boldsymbol u} \rho \mathcal{A} \right) &= \rho D_t \mathcal{A} + \mathcal{A} \left[ \partial_t \rho + \nabla \cdot \left( {\boldsymbol u} \rho \right) \right] \\
&= \rho D_t \mathcal{A}
\end{split}
\label{eqn:time_derivative_conversion}
\end{equation}
Eqs. \ref{eqn:continuity_noise_averaged}, \ref{eqn:momentum_noise_averaged2}, and \ref{eqn:spin_noise_averaged2} can then be written as
\begin{equation}
D_t \rho = - \rho \nabla \cdot {\boldsymbol u}
\label{eqn:avg_continuity_generic}
\end{equation}
\begin{equation}
\rho D_t {\boldsymbol u} = - \frac{\gamma}{m} \rho {\boldsymbol u} + \nabla \cdot \tau
\label{eqn:avg_linear_momentum_generic}
\end{equation}
and
\begin{equation}
\rho D_t \mathcal{L} = - \frac{\gamma}{m} \rho \mathcal{L} + \frac{F_a d_0}{2 m} \rho + 2 p^* + \nabla \cdot {\boldsymbol c}
\label{eqn:avg_spin_generic}
\end{equation}

We need to make one final adjustment to Eq. \ref{eqn:avg_spin_generic}. As written, this equation contains two distinct contributions to the spin density: that from the unforced nonequilibrium steady state, which is set by the balance of spin injection, antisymmetric stress, and bath dissipation, and the additional spin density induced by external forcing and nonvanishing flow. In order to separate out these contributions, we will assume that, even in the presence of spontaneous boundary current formation, we can define a bulk region in the unforced nonequilibrium steady state in which the velocity vanishes and the density, antisymmetric stress, and spin density are constant and indicated by $\rho_0$, $p_0^*$, and $\mathcal{L}_0$, respectively. In this case, since the velocity vanishes and the spin density is uniform, the couple stress ${\boldsymbol c}$ must vanish and we have from Eq. \ref{eqn:avg_spin_generic}
\begin{equation}
0 = - \frac{\gamma}{m} \rho_0 \mathcal{L}_0 + \frac{F_a d_0}{2 m} \rho_0 + 2 p^*_0
\label{eqn:bulk_spin}
\end{equation}
Further, from Eqs. \ref{eqn:spin_interaction_generic_decomp} and \ref{eqn:source_term_spin_eqn}, we have
\begin{equation}
p_0^* = \frac{1}{2} \left\langle \sum_i \frac{1}{2} \varepsilon : \Delta {\boldsymbol F}_{c,i} \delta \left( {\boldsymbol x} - {\boldsymbol x}_i \right) \right\rangle_0
\label{eqn:p0star}
\end{equation}
where a subscript zero indicates an average in the unforced nonequilibrium steady state. We therefore learn that {\it the antisymmetric stress in the unforced nonequilibrium steady state indicates the average inter-dimer interaction retarding free dimer spin}. We may relate it directly to the deviation of $\mathcal{L}_0$ from the ideal spin $\mathcal{L}_{\rm id}$, obtained when the inter-dimer friction vanishes and spin injection is exactly compensated by bath dissipation:
\begin{equation}
0 = - \frac{\gamma}{m} \rho_0 \mathcal{L}_{\rm id} + \frac{F_a d_0}{2 m} \rho_0
\label{eqn:Lid}
\end{equation}
Hence, from Eqs. \ref{eqn:bulk_spin} and \ref{eqn:Lid}, we find
\begin{equation}
p_0^* = \frac{1}{2} \frac{\gamma}{m} \rho_0 \left( \mathcal{L}_0 - \mathcal{L}_{\rm id} \right) \leq 0
\label{eqn:pstar0_L}
\end{equation}
That is, $p_0^*$ exactly quantifies the deviation of the true background spin from the ideal value and is necessarily nonpositive.

Subtracting Eq. \ref{eqn:bulk_spin} from \ref{eqn:avg_spin_generic} gives our final result for the {\it spin anomaly} balance equation
\begin{equation}
\begin{split}
\rho D_t \Delta \mathcal{L} = - &\frac{\gamma}{m} \left( \rho \Delta \mathcal{L} + \Delta \rho \mathcal{L}_0 \right) + \frac{F_a d_0}{2 m} \Delta \rho \\
&+ 2 \left( p^* - p^*_0 \right) + \nabla \cdot {\boldsymbol c}
\end{split}
\label{eqn:avg_spin_anomaly_generic}
\end{equation}
where $\Delta \mathcal{L} \equiv \mathcal{L} - \mathcal{L}_0$ and $\Delta \rho \equiv \rho - \rho_0$. As it stands, the frictional term proportional to $\Delta \rho \mathcal{L}_0$ appearing in this equation still requires us to measure $\mathcal{L}_0$ in simulation. This term exactly vanishes in the incompressible limit in which $\Delta \rho = 0$.

\subsection{Id. Phenomenology of an isotropic, two-dimensional, chiral fluid}

In order to make any progress using the balance equations given by Eqs. \ref{eqn:avg_continuity_generic}, \ref{eqn:avg_linear_momentum_generic}, and \ref{eqn:avg_spin_anomaly_generic}, we need to determine the manner in which our microscopic fluxes $\tau$ and ${\boldsymbol c}$ are permitted to couple to the relevant kinematic fields --  namely, the velocity gradient $\nabla {\boldsymbol u}$, spin density anomaly $\Delta \mathcal{L}$, and spin density anomaly gradient $\nabla \Delta \mathcal{L}$. We will take an approach somewhat distinct from the usual ones \cite{Avron, Kranthi1, Banerjee, free_surface, waves_Kranthi}. Rather than determining the form permitted by isotropy and two-dimensionality of, for example, the general fourth-rank viscosity tensor linking the stress tensor to the velocity gradient, we will expand the microscopic fluxes and kinematic fields into all of their distinct components based on tensorial character (scalar, vector, or second rank tensor) and symmetry (symmetric and antisymmetric components of second rank tensors). This allows us to identify by eye the permitted couplings between components of the fluxes and kinematic fields and expand the fluxes in scalar response coefficients.

Before deriving out constitutive relations, we briefly recapitulate why the rheology of two-dimensional isotropic, Newtonian fluids is unique. The emergence of the odd viscosity in a two-dimensional chiral active fluid originates from the fact that the second rank Levi-Civita tensor is isotropic in two dimensions \cite{Avron}. This is in contrast to three dimensions, where the only isotropic second rank tensor is the identity. Hence, in what follows, we will have to allow for an antisymmetric contribution to the second rank stress tensor proportional to the Levi-Civita tensor, and we will have to consider vector and tensor duals -- obtained by contraction with the Levi-Civita tensor -- in elucidating the fundamental components of the kinematic fields to which the microscopic fluxes may couple. It is this latter consideration that will lead to the emergence of the odd viscosity, among other novel transport coefficients.

We begin by expanding the velocity gradient:
\begin{equation}
\begin{split}
\nabla {\boldsymbol u} &= \frac{1}{2} \left( \nabla {\boldsymbol u} + \nabla {\boldsymbol u}^T - \nabla \cdot {\boldsymbol u} \mathbbm{1} \right) + \frac{1}{2} \nabla \cdot {\boldsymbol u} \mathbbm{1} + \omega \varepsilon \\
&\equiv E + \frac{1}{2} \nabla \cdot {\boldsymbol u} \mathbbm{1} + \omega \varepsilon
\end{split}
\label{eqn:strain_decomp}
\end{equation}
Here, $E$ represents the symmetric, traceless part of the velocity gradient corresponding to irrotational, volume-preserving strain of a fluid element. The term proportional to the velocity divergence represents the component of the strain that modifies the volume of the fluid element, and the term $\omega \varepsilon$, with $\omega \equiv \varepsilon : \nabla {\boldsymbol u}$ the vorticity, is the antisymmetric part of the velocity tensor, representing uniform solid body rotation of a fluid element.

We therfore must consider coupling to the scalars $\nabla \cdot {\boldsymbol u}$ and $\omega$ and the symmetric, traceless second rank tensor $E$, in addition to the scalar $\Delta \mathcal{L}$ and vector $\nabla \Delta \mathcal{L}$. We must also include the second rank isotropic tensors formed by multiplication of the scalars $\nabla \cdot {\boldsymbol u}$, $\omega$, and $\Delta \mathcal{L}$ by the isotropic tensors $\mathbbm{1}$ and $\varepsilon$. However, this does not complete the list: because of the isotropy of the Levi-Civita tensor, we must also consider the duals $\varepsilon \cdot E$, $E \cdot \varepsilon$, and $\varepsilon \cdot \nabla \Delta \mathcal{L} \equiv \nabla^* \Delta \mathcal{L}$. Furthermore, the second rank tensors $\varepsilon \cdot E$ and $E \cdot \varepsilon$ must be properly symmetrized and antisymmetrized. Recognizing that $\left( \varepsilon \cdot E \right)^T = - E \cdot \varepsilon$, easily verifiable by direct calculation in Cartesian coordinates, we find that $\varepsilon \cdot E$ and $E \cdot \varepsilon$ combine into one symmetric tensor, $\varepsilon \cdot E - E \cdot \varepsilon$, and one antisymmetric tensor, $\varepsilon \cdot E + E \cdot \varepsilon$. Furthermore, one may show that $\varepsilon \cdot E + E \cdot \varepsilon = \left( \nabla \cdot {\boldsymbol u} \right) \varepsilon$, and its inclusion is therefore redundant. Thus, the list of fundamental kinematic variables becomes, in order of ascending rank: $\omega$, $\nabla \cdot {\boldsymbol u}$, $\Delta \mathcal{L}$; $\nabla \Delta \mathcal{L}$,  $\nabla^* \Delta \mathcal{L}$; $\omega \mathbbm{1}$, $\nabla \cdot {\boldsymbol u} \mathbbm{1}$, $\Delta \mathcal{L} \mathbbm{1}$, $\omega \varepsilon$, $\nabla \cdot {\boldsymbol u} \varepsilon$, $\Delta \mathcal{L} \varepsilon$, and $\varepsilon \cdot E - E \cdot \varepsilon$.

We now decompose the stress tensor in the same manner as the velocity gradient:
\begin{equation}
\tau = - p \mathbbm{1} + p^* \varepsilon + \sigma
\label{eqn:tau_decomp}
\end{equation}
with the mechanical pressure
\begin{equation}
p \equiv - \frac{1}{2} {\rm tr} \tau
\label{eqn:mechanical_pressure}
\end{equation}
the antisymmetric stress
\begin{equation}
p^* \varepsilon \equiv \frac{1}{2} \left( \tau - \tau^T \right)
\label{eqn:pstar_defn}
\end{equation}
and the deviatoric stress
\begin{equation}
\sigma \equiv \frac{1}{2} \left( \tau + \tau^T - {\rm tr} \, \tau \mathbbm{1} \right)
\label{eqn:deviatoric_stress}
\end{equation}
which is by construction symmetric and traceless.

Finally, expanding $p^*$ and $\sigma$ in those kinematic variables that are permitted by rank and symmetry gives
\begin{equation}
p^* = p^*_0 + \eta_R \omega - \phi \Delta \mathcal{L} + \lambda \nabla \cdot {\boldsymbol u}
\label{eqn:pstar_expansion}
\end{equation}
and
\begin{equation}
\sigma = 2 \eta_S E + \eta_O \left( \varepsilon \cdot E - E \cdot \varepsilon \right) \equiv \sigma^S + \sigma^O
\label{eqn:sigma_decomp}
\end{equation}
In Eq. \ref{eqn:pstar_expansion}, the transport coefficients are the rotational viscosity $\eta_R$, linking vorticity to the antisymmetric stress; $\phi$, which links the spin anomaly to the antisymmetric stress and, from Eq. \ref{eqn:spin_final} below, is best interpreted as quantifying the increase in inter-dimer friction which tends to efface $\Delta \mathcal{L}$ and return the spin density to its bulk value; and a coefficient $\lambda$ linking the divergence of the velocity field to the antisymmetric stress.

We have introduced in Eq. \ref{eqn:sigma_decomp} the familiar shear viscosity, as well as the odd viscosity.

We note that in the case of passive fluids the mechanical pressure, Eq. \ref{eqn:mechanical_pressure}, would be decomposed into some reference value $p_{\rm state}$, regarded as a thermodynamic state variable, and a deviation expanded in the kinematic variables. This would lead to a decomposition of the form $p = p_{\rm state} - \eta_B \nabla \cdot {\boldsymbol u} + c_1 \omega + c_2 \Delta \mathcal{L}$, introducing the familiar bulk viscosity $\eta_B$ as well as other transport coefficients governing the coupling to the vorticity and spin anomaly. However, the expansion point $p_{\rm state}$ is known to be ill-defined in active fluids \cite{active_pressure1, active_pressure2}, and we therefore regard the total mechanical pressure, cleanly defined as the negative one-half trace of the stress tensor, as a mechanical variable to be determined by the hydrodynamic equations of motion.

Finally, we may decompose the couple stress as
\begin{equation}
{\boldsymbol c} = \alpha_e \nabla \Delta \mathcal{L} + \alpha_o \nabla^* \Delta \mathcal{L}
\label{eqn:couple_stress}
\end{equation}
with $\alpha_e$ and $\alpha_o$ the even and odd spin diffusivities, respectively. Note that $\nabla \cdot {\boldsymbol c} = \alpha_e \nabla^2 {\boldsymbol c}$, and therefore $\alpha_o$ can only be dynamically relevant if it appears in a boundary condition. This is similar to the odd diffusivity of passive tracers in chiral active fluids \cite{Cory_odd_diffusivity, my_PRL}.


Inserting the definition of the volume-preserving strain rate tensor $E$ (Eq. \ref{eqn:strain_decomp}) into our phenomenological expression for $\sigma^O$ (Eq. \ref{eqn:sigma_decomp}) and taking the divergence, we obtain
\begin{equation}
\begin{split}
\nabla \cdot \sigma^O &= \frac{1}{2} \eta_O \left( \nabla^2 {\boldsymbol u}^* + \nabla \omega + \nabla^* \nabla \cdot {\boldsymbol u} \right) \\
&= \eta_O \nabla^2 {\boldsymbol u}^* + \frac{1}{2} \eta_O \left( \nabla \omega - \nabla^2 {\boldsymbol u}^* \right) + \frac{1}{2} \eta_O \nabla^* \nabla \cdot {\boldsymbol u}
\label{eqn:DsigmaO_1}
\end{split}
\end{equation}
We can show that the second and third terms in the second equality exactly cancel by rewriting the second term as $- (\eta_O/2) \left( \nabla \otimes \nabla^* + \varepsilon \nabla^2 \right) \cdot {\boldsymbol u}$ and using the identity $\nabla \otimes \nabla^* + \varepsilon \nabla^2 = \nabla^* \otimes \nabla$, easily verifiable by direct calculation in Cartesian coordinates. We then find the simple expression
\begin{equation}
\nabla \cdot \sigma^O = \eta_O \nabla^2 {\boldsymbol u}^*
\label{eqn:DsigmaO_final}
\end{equation}

\subsection{Ie. Compressible hydrodynamic equations}

Inserting Eqs. \ref{eqn:tau_decomp}, \ref{eqn:pstar_expansion}, \ref{eqn:sigma_decomp}, \ref{eqn:couple_stress}, and \ref{eqn:DsigmaO_final} into Eqs. \ref{eqn:avg_continuity_generic}, \ref{eqn:avg_linear_momentum_generic}, and \ref{eqn:avg_spin_anomaly_generic}, we finally obtain for our general hydrodynamic equations
\begin{equation}
D_t \rho = - \rho \nabla \cdot {\boldsymbol u}
\label{eqn:continuity_final}
\end{equation}
\begin{equation}
\begin{split}
\rho D_t {\boldsymbol u} = &- \frac{\gamma}{m} \rho {\boldsymbol u} - \nabla p + \left( \eta_S + \eta_R \right) \nabla^2 {\boldsymbol u} + \eta_O \nabla^2 {\boldsymbol u}^* \\
&+ \phi \nabla^* \Delta \mathcal{L} - \left( \eta_R \nabla + \lambda \nabla^* \right) \nabla \cdot {\boldsymbol u}
\end{split}
\label{eqn:momentum_final}
\end{equation}
and
\begin{equation}
\begin{split}
\rho D_t \Delta \mathcal{L} = &- \frac{\gamma}{m} \left( \rho \Delta \mathcal{L} + \Delta \rho \mathcal{L}_0 \right) + \frac{F_a d_0}{2m} \Delta \rho \\
&+ 2 \left( \eta_R \omega - \phi \Delta \mathcal{L} + \lambda \nabla \cdot {\boldsymbol u} \right) + \alpha_e \nabla^2 \Delta \mathcal{L}
\end{split}
\label{eqn:spin_final}
\end{equation}
Interestingly, Eq. \ref{eqn:momentum_final} reveals that the effect of the rotational viscosity on the balance of linear momentum is to enhance the rate of linear momentum diffusion through the term $\eta_R \nabla^2 {\boldsymbol u}$ and to induce a dependence on the velocity divergence through the term $- \eta_R \nabla \nabla \cdot {\boldsymbol u}$, similar to the traditional bulk viscosity but with opposite sign.

\subsection{If. Hydrodynamic equations in the incompressible limit}

The incompressible forms of Eqs. \ref{eqn:continuity_final} through \ref{eqn:spin_final} are given by
\begin{equation}
\nabla \cdot {\boldsymbol u} = 0
\label{eqn:continuity_incompressible}
\end{equation}
\begin{equation}
\begin{split}
\rho_0 D_t {\boldsymbol u} = &- \frac{\gamma}{m} \rho_0 {\boldsymbol u} - \nabla p + \left( \eta_S + \eta_R \right) \nabla^2 {\boldsymbol u} + \eta_O \nabla^2 {\boldsymbol u}^* \\
&+ \phi \nabla^* \Delta \mathcal{L}
\end{split}
\label{eqn:momentum_incompressible}
\end{equation}
and
\begin{equation}
\rho_0 D_t \Delta \mathcal{L} = - \frac{\gamma}{m} \rho_0 \Delta \mathcal{L} + \alpha_e \nabla^2 \Delta \mathcal{L} + 2 \left( \eta_R \omega - \phi \Delta \mathcal{L} \right)
\label{eqn:Deltaspin_incompressible}
\end{equation}
We see that in this limit Eq. \ref{eqn:Deltaspin_incompressible} represents an independent balance for the spin anomaly $\Delta \mathcal{L}$, entirely decoupled from the bulk value $\mathcal{L}_0$.

\subsection{Ig. Comparison to previous results}

It is instructive to compare our general method for the derivation of hydrodynamic equations in active systems and the particular results we have obtained for our chiral active fluid model to previous works on the subject. Ref. \cite{joanny} is the classic reference on active hydrodynamics. In their review, the authors focus primarily on phenomenological derivations of hydrodynamic equations for active systems, citing the lack of a simple, universal method for deriving hydrodynamic equations from microscopic models -- though they do outline an approximate procedure for microscopic course graining from the Smoluchowski equation for an overdamped system. They assume that all systems in which dry friction is present are necessarily overdamped. Our method is both simpler and exact, in that it follows directly from the stochastic equations of motion, and more general, in that we allow for the case of low friction systems in which neither substrate friction nor inertia is negligible. Indeed, our procedure could in principle allow for more complex substrate frictions accounting for, {\it e.g.}, flows induced in an underlying fluid bath by introducing a non-Markovian frictional kernel in a generalized Langevin equation.

Ref. \cite{Vitellis_pretty_pictures} derives fluctuating hydrodynamic equations for a chiral active fluid through the introduction of an effective temperature and the extension of the Mori-Zwanzig procedure to a system far from equilibrium. Our method does not assume an effective temperature description and does not require the extension of near-equilibrium procedures. Ref. \cite{Lubensky_PB} uses a field-theoretic Poisson bracket formalism to derive dissipative hydrodynamic equations from a coarse-grained Hamiltonian and a phenomenological dissipation kernel. Our method does not require such phenomenological considerations, though it does require the introduction of a bath temperature, unlike the method of Ref. \cite{Lubensky_PB}, which coarse grains only spatio-temporally.

Ref. \cite{Kranthi_hydro_eqns, Kranthi1, Cory_MD} consider the same model that we do here. Their coarse graining procedure is based on the introduction of a smooth coarse graining function, rather than a delta function, by which they define the hydrodynamic densities. Their procedure is similar to our own, with a few key differences. They define their densities in terms of the monomer coordinates, rather than partitioning into dimer center of mass and spin degrees of freedom, and their results necessarily conflate the linear and spin angular momenta. They do not rewrite the noise in terms of hydrodynamic variables as we do here, and therefore their fluctuating hydrodynamic equations are not closed -- though this becomes irrelevant when the equations are noise-averaged. Finally, the stress tensor they obtain is defined in terms of monomer, rather than dimer, coordinates. In particular, this leads to an expression for the magnitude of the antisymmetric stress in the steady state $p^*_0 = - F_a d_0 \rho_0 / 4 m$, which would suggest that the torque induced on a fluid element in the bulk steady state $2 p_0^*$ exactly compensates the spin injection $F_a d_0 \rho_0 / 2m$, and, from their spin balance equation, that the total spin in the steady state therefore exactly vanishes. For these reasons, we believe our expression for and interpretation of the stress tensor and antisymmetric stress -- as the quantity which characterizes the frustration of free particle spin in the steady state -- is the correct one.

\section{II. Boundary current solution}

\subsection{IIa. Angular momentum balance}

From our mesoscopic equations of motion (Eqs. \ref{eqn:xeom_monomer} and \ref{eqn:peom_monomer}), we can derive the equations of motion for the noise-averaged total angular momentum $J \equiv \left\langle \sum_{i \alpha} \varepsilon : {\boldsymbol x}_i^{\alpha} \otimes {\boldsymbol p}_i^{\alpha} \right\rangle$, spin angular momentum $L \equiv \left\langle \sum_i \varepsilon : \left( {\boldsymbol d}_i / 2 \right) \otimes \Delta {\boldsymbol F}_i \right\rangle$, and orbital angular momentum $Q \equiv \left\langle \sum_i {\boldsymbol x}_i \otimes {\boldsymbol F}_i \right\rangle$, where here angled brackets denote a noise average not necessarily in the steady state. We have
\begin{equation}
\dot{J} = - \frac{\gamma}{m} J + N_{\rm dimer} F_a d_0
\label{eqn:J_balance}
\end{equation}
\begin{equation}
\dot{L} = - \frac{\gamma}{m} L + N_{\rm dimer} F_a d_0 + \left\langle \sum_i \varepsilon : \frac{1}{2} {\boldsymbol d}_i \otimes \Delta {\boldsymbol F}_i \right\rangle
\label{eqn:L_balance}
\end{equation}
and
\begin{equation}
\dot{Q} = - \frac{\gamma}{m} Q + \left\langle \sum_i \varepsilon : {\boldsymbol x}_i \otimes {\boldsymbol F}_i \right\rangle
\label{eqn:Q_balance}
\end{equation}
We have neglected any interaction of the particles with the frictionless boundary. While the boundary cannot change the global value of the total angular momentum because it is frictionless and circular, it may frustrate the local free spin of a rotor close to the boundary, and correspondingly increase the orbital angular momentum at the boundary such that its total contribution to $J$ vanishes. This is a negligible correction that does not affect our conclusions.

The sum of the final terms in Eqs. \ref{eqn:L_balance} and \ref{eqn:Q_balance} is given by
\begin{equation}
\begin{split}
&\left\langle \sum_i \varepsilon : \frac{1}{2} {\boldsymbol d}_i \otimes \Delta {\boldsymbol F}_i \right\rangle + \left\langle \sum_i \varepsilon : {\boldsymbol x}_i \otimes {\boldsymbol F}_i \right\rangle \\
&\quad\quad\quad\quad= \left\langle \sum_{i\alpha} {\boldsymbol x}_i^{\alpha} \otimes {\boldsymbol F}_i^{\alpha} \right\rangle = 0
\end{split}
\label{eqn:interaction_sum}
\end{equation}
-- that is, by the sum of all internal torques in the system, which must necessarily vanish given the central nature of these interactions. This is a key difference between our model and other colloidal models \cite{free_surface, Vitellis_pretty_pictures}, which allow for non-central interactions between colloidal particles. In this case, the interaction terms appearing in Eqs. \ref{eqn:L_balance} and \ref{eqn:Q_balance} would not exactly compensate, representing the ability of inter-rotor collisions to generate friction and implicitly dissipate energy to the environment in the absence of an explicit environmental coupling ({\it e.g.}, substrate friction).

From Eq. \ref{eqn:interaction_sum} and the area integral of Eq. \ref{eqn:avg_spin_generic} over the system, we can relate the interaction terms to the antisymmetric stress $p^*$ by
\begin{equation}
\begin{split}
2 \int_A {\rm d}A \, p^* \equiv 2 A p_A^* &= \left\langle \sum_i \varepsilon : \frac{1}{2} {\boldsymbol d}_i \otimes \Delta {\boldsymbol F}_i \right\rangle \\
&= - \left\langle \sum_i \varepsilon : {\boldsymbol x}_i \otimes {\boldsymbol F}_i \right\rangle
\end{split}
\label{eqn:Tint}
\end{equation}
where a subscript $A$ denotes an area average. This allows us to write Eqs. \ref{eqn:J_balance} through \ref{eqn:Q_balance} as
\begin{equation}
\dot{J} = - \frac{\gamma}{m} J + N_{\rm dimer} F_a d_0
\label{eqn:J_dimer_2}
\end{equation}
\begin{equation}
\dot{L} = - \frac{\gamma}{m} L + N_{\rm dimer} F_a d_0 + 2 A p_A^*
\label{eqn:L_dimer_2}
\end{equation}
and
\begin{equation}
\dot{Q} = - \frac{\gamma}{m} Q - 2 A p_A^*
\label{eqn:Q_dimer_2}
\end{equation}
When $\gamma \neq 0$, we can solve these equations in the steady state obtained for arbitrary initial conditions as $t \rightarrow \infty$ to obtain
\begin{equation}
J_{\infty} = \frac{m}{\gamma} N_{\rm dimer} F_a d_0
\label{eqn:J_steadystate_dimer}
\end{equation}
\begin{equation}
L_{\infty} = \frac{m}{\gamma} \left( N_{\rm dimer} F_a d_0 + 2 A p_{A,\infty}^* \right)
\label{eqn:L_steadystate_dimer}
\end{equation}
and
\begin{equation}
Q_{\infty} = - \frac{m}{\gamma} 2 A p_{A,\infty}^*
\label{eqn:Q_steadystate_dimer}
\end{equation}
These results reveal the physics of boundary current formation in our system. Collisions between dimers lead to a frustration of free spin at finite density and a nonzero value of the antisymmetric stress. Because our interactions are central, these collisions cannot dissipate angular momentum or rotational energy; they can only transfer energy from the spin to the orbital degree of freedom. At steady state, conservation of the total angular momentum, set by the balance between angular momentum injection $N_{\rm dimer} F_a d_0$ and the total substrate angular momentum dissipation $(\gamma/m) J_{\infty}$, requires that a boundary current carrying orbital angular momentum form to compensate the decrease in spin due to inter-dimer interactions. This transfer from spin to orbital angular momentum is mediated by $p^*$.

Our model supports no steady state in the absence of substrate friction ({\it i.e.}, when $\gamma = 0$). While a steady value of $L$ may possibly be obtained from Eq. \ref{eqn:L_dimer_2} for $\gamma = 0$ when $N_{\rm dimer} F_a d_0 = - 2 A p^*_{A,\infty}$, Eqs. \ref{eqn:Q_dimer_2} and \ref{eqn:J_dimer_2} show that the orbital, and therefore the total, angular momentum will necessarily diverge as $t \rightarrow \infty$. This is in contrast to colloidal models with non-central interactions; the frictional forces developed during inter-rotor collisions implicitly dissipates energy and allows for the possibility of a steady state even in the absence of substrate friction. Even in this case, the mechanism of boundary current formation is fundamentally the same. Inter-rotor friction introduces an additional mechanism to dissipate the angular momentum injected into the system, but the frustration of free spin at finite density still means that some portion of the remaining total angular momentum must be transferred from spin to orbital angular momentum, requiring the formation of a boundary current.

\subsection{IIb. Dual vectors in cylindrical coordinates}

In any geometry and coordinate system, the dual of a vector ${\boldsymbol v}$ is defined generically as the inner product of the Levi-Civita {\it tensor} $\tilde{\varepsilon}$ with that vector: ${\boldsymbol v}^* \equiv \tilde{\varepsilon} \cdot {\boldsymbol v}$. In Cartesian coordinates, this definition is unambiguous because $\tilde{\varepsilon}$ coincides with the {\it antisymmetric symbol} $\varepsilon$, defined for any given basis in any geometry and coordinate system as either the sign of the permutation of its indices necessary to recover the ordered set in the case that no indices are repeated or zero otherwise -- {\it e.g.}, in two-dimensions $\varepsilon_{12} = - \varepsilon_{21} = +1$ and $\varepsilon_{11} = \varepsilon_{22} = 0$. This symbol does not transform as a tensor, however, and it therefore must be modified for generic coordinate systems. This leads to the introduction of the Levi-Civita tensor $\tilde{\varepsilon}$, which is defined uniquely by the requirement that it transform as a tensor and the condition that $\tilde{\varepsilon} = \varepsilon$ in Cartesian coordinates. Because this tensor necessarily appears in the equations governing the flow of an odd viscous fluid through the introduction of vector duals, and because most of the following theoretical development will be conducted in cylindrical coordinates, we pause in this section to derive the proper form of the Levi-Civita tensor and vector dual in cylindrical coordinates.

We will start by writing Eqs. \ref{eqn:continuity_final} through \ref{eqn:spin_final} in component notation for a generic coordinate system. Before doing that, however, we note that in this section we will (out of necessity) distinguish between contravariant and covariant indices \cite{Landau_Vol2} by whether they appear as superscripts or subscripts, respectively, and we will adopt Einstein summation notation, in which a repeated index, appearing once as a contravariant index and once as a covariant index, is understood to imply summation over that index -- {\it e.g.}, $a_i b^i \equiv \sum_i a_i b^i$. Further, we recall that raising and lowering of indices is accomplished by contraction with the metric tensor $g_{ij}$ -- defined through the relation $d \ell^2 \equiv g_{ij} dx^i dx^j$ for an infinitesimal displacement $d \ell$ and an arbitrary coordinate system $\left\{ x^i \right\}$ -- or its inverse $g^{-1}_{ij} \equiv g^{ij}$. With these points in mind, we write Eqs. \ref{eqn:continuity_final} through \ref{eqn:spin_final} as
\begin{equation}
\left( \partial_t + u^i \nabla_i \right) \rho = - \rho \nabla_i u^i
\label{eqn:continuity_component_generic}
\end{equation}
\begin{equation}
\begin{split}
\rho &\left( \partial_t + u^j \nabla_j \right) u^i = - \frac{\gamma}{m} \rho u^i - g^{ij} \nabla_j p \\
&+ \left( \eta_S + \eta_R \right) g^{jk} \nabla_j \nabla_k u^i + \eta_O g^{jk} \nabla_j \nabla_k g^{il} \tilde{\varepsilon}_{lm} u^m \\
&+ \phi \tilde{\varepsilon}^{ij} \nabla_j \Delta \mathcal{L} - \left( \eta_R g^{ij} \nabla_j + \lambda \tilde{\varepsilon}^{ij} \nabla_j \right) \nabla_k u^k
\end{split}
\label{eqn:lin_momentum_component_generic}
\end{equation}
and
\begin{equation}
\begin{split}
\rho &\left( \partial_t + u^i \nabla_i \right) \Delta \mathcal{L} = - \frac{\gamma}{m} \left( \rho \Delta \mathcal{L} + \Delta \rho \mathcal{L}_0 \right) + \frac{F_a d_0}{2m} \Delta \rho \\
&+ 2 \left( \eta_R \tilde{\varepsilon}_{ij} g^{ik} \nabla_k u^j - \phi \Delta \mathcal{L} + \lambda \nabla_i u^i \right) \\
&+ \alpha_e g^{ij} \nabla_i \nabla_j \Delta \mathcal{L}
\end{split}
\label{eqn:spin_component_generic}
\end{equation}
where $\nabla_i$ indicates the $i$-th component of the covariant derivative rather than the ordinary gradient (though they coincide in Cartesian coordinates). See, {\it e.g.}, Ref. \cite{Landau_Vol2} for a relatively concise discussion. Examining Eqs. \ref{eqn:continuity_component_generic} through \ref{eqn:spin_component_generic}, we see that we will be concerned with two forms of the vector dual, both appearing in Eq. \ref{eqn:lin_momentum_component_generic}: $u^{*i} \equiv g^{il} \tilde{\varepsilon}_{lm} u^m$ and $\nabla^{*i} \equiv \tilde{\varepsilon}^{ij} \nabla_j$. We recognize that the latter term, $\nabla^{*i}$, acts on a scalar both times it appears in Eq. \ref{eqn:lin_momentum_component_generic}, either the spin density anomaly $\Delta \mathcal{L}$ or the trace of the velocity gradient $\nabla_k u^k$. The components of the covariant derivative $\nabla_i$ are therefore in this case simply the components of the cylindrical gradient -- i.e., $\nabla = \hat{\boldsymbol r} \partial_r + \hat{\boldsymbol \theta} r^{-1} \partial_{\theta}$. The term $\omega \equiv \tilde{\varepsilon}_{ij} g^{ik} \nabla_k u^j$ proportional to the Levi-Civita tensor appearing in Eq. \ref{eqn:spin_component_generic} is simply the vorticity, the form of which in cylindrical coordinates is compiled in standard references and need not be derived here.

One may show from the condition that $\tilde{\varepsilon} = \varepsilon$ in Cartesian coordinates that the contravariant and covariant components of the Levi-Civita tensor $\tilde{\varepsilon}$ are related to the antisymmetric symbol $\varepsilon$ by the relations \cite{Landau_Vol2}
\begin{equation}
\tilde{\varepsilon}^{ij} = \frac{1}{\sqrt{g}} \varepsilon^{ij}; \quad \tilde{\varepsilon}_{ij} = \sqrt{g} \varepsilon_{ij}
\label{eqn:LCT}
\end{equation}
where $g \equiv {\rm det} \left( g_{ij} \right)$ is the determinant of the metric tensor. We then have for $u^{*i}$ and $\nabla^{*i}$
\begin{equation}
\nabla^{*i} \equiv \tilde{\varepsilon}^{ij} \nabla_j = \frac{1}{\sqrt{g}} \varepsilon^{ij} \nabla_j
\label{eqn:dual_gradient2}
\end{equation}
and
\begin{equation}
u^{*i} \equiv g^{ij} \tilde{\varepsilon}_{jk} u^k = \sqrt{g} g^{ij} \varepsilon_{jk} u^k
\label{eqn:dual_vector2}
\end{equation}

From the relation $d \ell^2 \equiv g_{ij} dx^i dx^j = dr^2 + r^2 d\theta^2$ we have for the metric tensor
\begin{equation}
\begin{split}
\left( g_{ij} \right) &=
\begin{pmatrix}
&1 &0 \\
&0 &r^2
\end{pmatrix} \\
&\Longrightarrow g = r^2, \quad \left( g_{ij} \right)^{-1} \equiv \left( g^{ij} \right) =
\begin{pmatrix}
&1 &0 \\
&0 &1/r^2
\end{pmatrix}
\end{split}
\label{eqn:metric_tensor_cylindrical_coordinates}
\end{equation}
Note that these matrices are written in the `natural' direct basis $\left\{ {\boldsymbol a}_i \right\}$ defined by $d {\boldsymbol \ell} = {\boldsymbol a}_i dx^i = \hat{\boldsymbol r} dr + r \hat{\boldsymbol \theta} d\theta$, from which we infer
\begin{equation}
{\boldsymbol a}_r = \hat{\boldsymbol r}; \quad {\boldsymbol a}_{\theta} = r \hat{\boldsymbol \theta}
\label{eqn:covariant_basis}
\end{equation}
Furthermore, by multiplying by the inverse metric tensor (Eq. \ref{eqn:metric_tensor_cylindrical_coordinates}), we obtain for the dual basis
\begin{equation}
{\boldsymbol a}^r = \hat{\boldsymbol r}; \quad {\boldsymbol a}^{\theta} = \frac{\hat{\boldsymbol \theta}}{r}
\label{eqn:contravariant_basis}
\end{equation}
Using these identities and Eq. \ref{eqn:metric_tensor_cylindrical_coordinates}, we may write the term $\sqrt{g} g^{ij}$ appearing in Eq. \ref{eqn:dual_vector2} in terms of the $\left\{ \hat{\boldsymbol r}, \hat{\boldsymbol \theta} \right\}$ basis as
\begin{equation}
\begin{split}
\sqrt{g} g^{ij} {\boldsymbol a}_i \otimes {\boldsymbol a}_j &= r \left( {\boldsymbol a}_r \otimes {\boldsymbol a}_r + \frac{1}{r^2} {\boldsymbol a}_{\theta} \otimes {\boldsymbol a}_{\theta} \right) \\
&= r \left( \hat{\boldsymbol r} \otimes \hat{\boldsymbol r} + \hat{\boldsymbol \theta} \otimes \hat{\boldsymbol \theta} \right)
\end{split}
\label{eqn:ggij2}
\end{equation}
Similarly, the antisymmetric symbols are given by
\begin{equation}
\begin{split}
\varepsilon_{ij} {\boldsymbol a}^i \otimes {\boldsymbol a}^j &\equiv {\boldsymbol a}^r \otimes {\boldsymbol a}^{\theta} - {\boldsymbol a}^{\theta} \otimes {\boldsymbol a}^r \\
&= \frac{1}{r} \left( \hat{\boldsymbol r} \otimes \hat{\boldsymbol \theta} - \hat{\boldsymbol \theta} \otimes \hat{\boldsymbol r} \right)
\end{split}
\label{eqn:eij1}
\end{equation}
and
\begin{equation}
\begin{split}
\varepsilon^{ij} {\boldsymbol a}_i \otimes {\boldsymbol a}_j &\equiv {\boldsymbol a}_r \otimes {\boldsymbol a}_{\theta} - {\boldsymbol a}_{\theta} \otimes {\boldsymbol a}_r \\
&= r \left( \hat{\boldsymbol r} \otimes \hat{\boldsymbol \theta} - \hat{\boldsymbol \theta} \otimes \hat{\boldsymbol r} \right)
\end{split}
\label{eqn:eij2}
\end{equation}
Inserting these results into Eqs. \ref{eqn:dual_gradient2} and \ref{eqn:dual_vector2}, we find that the dual gradient and velocity are given simply by
\begin{equation}
\begin{split}
\nabla^* &= \left( \hat{\boldsymbol r} \otimes \hat{\boldsymbol \theta} - \hat{\boldsymbol \theta} \otimes \hat{\boldsymbol r} \right) \cdot \nabla \\
&= \frac{\partial_{\theta}}{r} \hat{\boldsymbol r} - \partial_r \hat{\boldsymbol \theta}
\end{split}
\label{eqn:dual_gradient_final}
\end{equation}
and
\begin{equation}
\begin{split}
{\boldsymbol u}^* &= \left( \hat{\boldsymbol r} \otimes \hat{\boldsymbol \theta} - \hat{\boldsymbol \theta} \otimes \hat{\boldsymbol r} \right) \cdot {\boldsymbol u} \\
&= v \hat{\boldsymbol r} - u \hat{\boldsymbol \theta}
\end{split}
\label{eqn:dual_velcoity_final}
\end{equation}
where $u$ and $v$ are, respectively, the $r$ and $\theta$ components of the velocity vector: ${\boldsymbol u} = u \hat{\boldsymbol r} + v \hat{\boldsymbol \theta}$.

\subsection{IIc. Boundary current solution}

We now solve Eqs. \ref{eqn:continuity_final} through \ref{eqn:spin_final} for the hydrodynamic structure of the boundary current, subject to the integral statement of angular momentum conservation
\begin{equation}
S_{\eta_O} \left| \left\langle Q \right\rangle_0 \right| = \int_0^R 2 \pi r {\rm d} r \, \rho_0 r v_0
\label{eqn:orbital}
\end{equation}
where $Q$ is the total {\it orbital} angular momentum associated with the motions of the dimer centers of mass -- a quantity that must be measured in simulation -- angled brackets denote a nonequilibrium ensemble average, a subscript zero denotes an average obtained in the unforced nonequilibrium steady state, $v_0$ is the tangential ($\theta$) component of the velocity, and $R$ is the radius of the circular confinement. Further, we have defined $S_{\eta_O} \equiv {\rm sign} \left( \eta_O \right)$ and noted that the sign of the odd viscosity coincides with the sign of the orbital angular momentum and the tangential velocity -- {\it e.g.}, $\eta_O > 0$ indicates counterclockwise dimer rotation, which must lead to counterclockwise ({\it i.e.}, positive) total orbital angular momentum and tangential boundary current velocity.

Rotational symmetry about the center of the circular confinement requires that, at steady state, the solution be independent of angular coordinate $\theta$ and the only nonvanishing component of the velocity be the tangential component. That is
\begin{equation}
{\boldsymbol u}_0 = v_0 (r) \hat{\boldsymbol \theta} \Longrightarrow {\boldsymbol u}_0^* = v_0 (r) \hat{\boldsymbol r}; \quad \partial_t = \partial_{\theta} = 0
\label{eqn:boundary_current_symmetry}
\end{equation}
where $r$ measures the radial distance from the center of the circular confinement, and the form of ${\boldsymbol u}_0^*$ follows from Eq. \ref{eqn:dual_velcoity_final}. This symmetry will greatly simplify Eqs. \ref{eqn:continuity_final} through \ref{eqn:spin_final}. In particular, this form of the velocity field is inherently solenoidal: $\nabla \cdot {\boldsymbol u}_0 = 0$. Furthermore, though this is not strictly required by the solenoidal condition, we observe in all of our simulations that $\rho_0$ is constant within the hydrodynamic region. With these simplifications, our governing equations reduce to
\begin{equation}
- \rho_0 \frac{v_0^2}{r} = - \partial_r p_0 + \eta_O \left( \partial_r^2 + \frac{\partial_r}{r} - \frac{1}{r^2} \right) v_0
\label{eqn:umomentum_boundary_current}
\end{equation}
\begin{equation}
0 = - \frac{\gamma}{m} \rho_0 v_0 + \left( \eta_S + \eta_R \right) \left( \partial_r^2 + \frac{\partial_r}{r} - \frac{1}{r^2} \right) v_0 - \phi \partial_r \Delta \mathcal{L}_0
\label{eqn:vmomentum_boundary_current}
\end{equation}
and
\begin{equation}
\begin{split}
0 = - \frac{\gamma}{m} \rho_0 \Delta \mathcal{L}_0 &+ \alpha_e \frac{\partial_r \left( r \partial_r \Delta \mathcal{L}_0 \right)}{r} \\
&+ 2 \left[ \eta_R \frac{\partial_r \left( r v_0 \right)}{r} - \phi \Delta \mathcal{L}_0 \right]
\end{split}
\label{eqn:spin_boundary_current}
\end{equation}
Note that the form of the $-\phi \partial_r \Delta \mathcal{L}_0$ term in Eq. \ref{eqn:vmomentum_boundary_current} follows from Eq. \ref{eqn:dual_gradient_final}. Before analyzing these equations, we rewrite them by solving Eq. \ref{eqn:umomentum_boundary_current} for $\left( \partial_r^2 + r^{-1} \partial_r - r^{-2} \right) v_0$, inserting the result into Eq. \ref{eqn:vmomentum_boundary_current}, and dividing through Eqs. \ref{eqn:vmomentum_boundary_current} and \ref{eqn:spin_boundary_current} by $\eta_S + \eta_R$. This gives
\begin{equation}
- \rho_0 \frac{v_0^2}{r} = - \partial_r p_0 + \eta_O \left( \kappa_{\gamma}^2 v_0 + \kappa_{\phi}^2 \partial_r \Delta \mathcal{L}_0 \right)
\label{eqn:pressure_eqn_boundary_current}
\end{equation}
\begin{equation}
\kappa_{\gamma}^2 v_0 + \kappa_{\phi}^2 \partial_r \Delta \mathcal{L}_0 = \left( \partial_r^2 + \frac{\partial_r}{r} - \frac{1}{r^2} \right) v_0
\label{eqn:velocity_eqn_boundary_current}
\end{equation}
and
\begin{equation}
\left( \kappa_{\gamma}^2 + 2 \kappa_{\phi}^2 \right) \Delta \mathcal{L}_0 - \mathcal{A}_e \frac{\partial_r \left( r \partial_r \Delta \mathcal{L}_0 \right)}{r} = 2 \mathcal{N}_R \frac{\partial_r \left( r v_0 \right)}{r}
\label{eqn:spin_eqn_boundary_current}
\end{equation}
where we have defined inverse length scales $\kappa_{\gamma}$ and $\kappa_{\phi}$ by
\begin{equation}
\kappa_{\gamma}^2 \equiv \frac{\gamma}{m} \frac{\rho_0}{\eta_S + \eta_R}; \quad \kappa_{\phi}^2 \equiv \frac{\phi}{\eta_S + \eta_R}
\label{eqn:friction_lengths}
\end{equation}
as well as the dimensionless parameters
\begin{equation}
\mathcal{A}_e \equiv \frac{\alpha_e}{\eta_S + \eta_R}; \quad \mathcal{N}_R \equiv \frac{\eta_R}{\eta_S + \eta_R}
\label{eqn:dim_params}
\end{equation}
which quantify, respectively, the relative importance of spin diffusivity and rotational viscosity compared to the effective linear momentum diffusivity $\eta_S + \eta_R$. The dimensionless parameter space will also contain the ratio of the inverse length scales defined in Eq. \ref{eqn:friction_lengths},
\begin{equation}
\frac{\kappa_{\phi}}{\kappa_{\gamma}} \equiv \sqrt{ \frac{m \phi}{\gamma \rho_0} }
\label{eqn:length_ratio}
\end{equation}
which characterizes the relative importance of $\phi$ compared to the linear bath friction $\gamma$, and a dimensionless scale of the system size $R$, which we take to be
\begin{equation}
\kappa_{\gamma} R \equiv \sqrt{ \frac{\gamma}{m} \frac{\rho_0}{\eta_S + \eta_R} } R
\label{eqn:dim_system_size}
\end{equation}
We take $\kappa_{\gamma}^{-1}$ as the relevant length scale because, as we will see in what follows, it typically characterizes the lateral thickness of the boundary current. Physically, this length scale is a measure of the radial distance over which the boundary current must extend so that shear stress is exactly balanced by bath friction.

In order to solve Eqs. \ref{eqn:pressure_eqn_boundary_current} through \ref{eqn:spin_eqn_boundary_current}, we will take the limit that $\kappa_{\gamma} R \gg 1$ so that we may neglect terms proportional to $r^{-1} \sim R^{-1}$ in comparison to radial gradients $\partial_r \sim \kappa_{\gamma}$. As indicated in main text Fig. 1, the exponential profiles obtained in this limit describe the observed profiles of velocity and spin excellently, indicating that this approximation is valid. In this regime, Eqs. \ref{eqn:pressure_eqn_boundary_current} through \ref{eqn:spin_eqn_boundary_current} simplify to
\begin{equation}
0 = - \partial_r p_0 + \eta_O \left( \kappa_{\gamma}^2 v_0 + \kappa_{\phi}^2 \partial_r \Delta \mathcal{L}_0 \right)
\label{eqn:pressure_eqn_boundary_current_final}
\end{equation}
\begin{equation}
\kappa_{\gamma}^2 v_0 + \kappa_{\phi}^2 \partial_r \Delta \mathcal{L}_0 = \partial_r^2 v_0
\label{eqn:velocity_eqn_boundary_current_final}
\end{equation}
and
\begin{equation}
\left( \kappa_{\gamma}^2 + 2 \kappa_{\phi}^2 \right) \Delta \mathcal{L}_0 - \mathcal{A}_e \partial_r^2 \Delta \mathcal{L}_0 = 2 \mathcal{N}_R \partial_r v_0
\label{eqn:spin_eqn_boundary_current_final}
\end{equation}
The boundary condition given in Eq. \ref{eqn:orbital} likewise simplifies in this limit:
\begin{equation}
S_{\eta_O} \left| \left\langle Q \right\rangle_0 \right| = 2 \pi R \rho_0 \int_0^{R} {\rm d} r \, r v_0
\label{eqn:orbital_simple}
\end{equation}

We search for solutions of Eqs. \ref{eqn:pressure_eqn_boundary_current_final} through \ref{eqn:spin_eqn_boundary_current_final} of the form
\begin{equation}
\begin{Bmatrix}
p_0 \\
v_0 \\
\Delta \mathcal{L}_0
\end{Bmatrix} =
\begin{Bmatrix}
P \\
V \\
L
\end{Bmatrix} e^{- K \left( R - r \right)}
\label{eqn:exponential_form_boundary_current}
\end{equation}
from which we find
\begin{equation}
P = \eta_O K V
\label{eqn:P_general}
\end{equation}
\begin{equation}
L = 2 \mathcal{N}_R \frac{K/\kappa_{\gamma}}{1 + 2 \left( \kappa_{\phi} / \kappa_{\gamma} \right)^2 - \mathcal{A}_e \left( K / \kappa_{\gamma} \right)^2} \kappa_{\gamma}^{-1} V
\label{eqn:L_general}
\end{equation}
and
\begin{equation}
\left( \frac{K}{\kappa_{\gamma}} \right)^2 - 1 = 2 \mathcal{N}_R \frac{\left( \kappa_{\phi} / \kappa_{\gamma} \right)^2}{1 + 2 \left( \kappa_{\phi} / \kappa_{\gamma} \right)^2 - \mathcal{A}_e \left( K / \kappa_{\gamma} \right)^2} \left( \frac{K}{\kappa_{\gamma}} \right)^2
\label{eqn:K_general}
\end{equation}
The value of $V$ is determined by the integral condition Eq. \ref{eqn:orbital_simple} and given by
\begin{equation}
V = S_{\eta_O} \frac{K}{2 \rho_0} \frac{\left| \left\langle Q \right\rangle_0 \right|}{\pi R^2}
\label{eqn:V_general}
\end{equation}
in the limit that $\kappa_{\gamma} R \rightarrow \infty$.

We will explore the solution for the velocity and spin profiles given by Eqs. \ref{eqn:L_general} through \ref{eqn:V_general} in different limiting regions of parameter space in the subsections that follow. Our simulation results (main text Figs. 1 and 2), and in particular the observation that we are able to extract accurate estimates of both $\eta_S$ and $\eta_R$ from observations of the boundary current, will determine which region of parameter space we are in.

\subsubsection{IIci. Limiting case I: $\mathcal{A}_e \rightarrow 0$}

We first examine the case that $\mathcal{A}_e \longrightarrow 0$ -- {\it i.e.}, the spin diffusivity is negligible relative to the linear momentum diffusivity, a scenario that seems intuitively likely from a microscopic view if we grant that inter-dimer collisions more easily transfer linear momentum between their centers of mass than spin momentum. In this limit, the quartic equation for the inverse decay length $K$, Eq. \ref{eqn:K_general}, reduces to a quadratic, the positive root of which is given by
\begin{equation}
\frac{K}{\kappa_{\gamma}} = \sqrt{\frac{1 + 2 \left( \kappa_{\phi} / \kappa_{\gamma} \right)^2}{1 + 2 \left( 1 - \mathcal{N}_R \right) \left( \kappa_{\phi} / \kappa_{\gamma} \right)^2}}
\label{eqn:K_LCI}
\end{equation}
The value of $v_0$ at the boundary, $V$, is unchanged and given by Eq. \ref{eqn:V_general}, while the boundary value of the spin anomaly $\Delta \mathcal{L}_0$ becomes
\begin{equation}
L = 2 \mathcal{N}_R \frac{K/\kappa_{\gamma}}{1 + 2 \left( \kappa_{\phi} / \kappa_{\gamma} \right)^2} \kappa_{\gamma}^{-1} V
\label{eqn:L_LCI}
\end{equation}
At this stage, the parametric dependencies of $K$ and $L$ are still too complex to allow us to extract estimates of the shear and rotational viscosities.

\subsubsection{IIcii. Limiting case II: $\mathcal{A}_e, \, \mathcal{N}_R \rightarrow 0$}

However, we note that, in the limit that $\mathcal{N}_R \rightarrow 0$, Eq. \ref{eqn:K_LCI} reduces to
\begin{equation}
K = \kappa_{\gamma}
\label{eqn:K_LCII}
\end{equation}
In this case, therefore, we may extract the $e$-folding length $\kappa_{\gamma}^{-1}$ from an exponential fit of the velocity profile and relate it to the sum $\eta_S + \eta_R$ via Eq. \ref{eqn:friction_lengths} and the specified parameters $\gamma$, $\rho_0$, and $m$. In this limit, $V$ reduces to
\begin{equation}
V = S_{\eta_O} \frac{\kappa_{\gamma}}{2 \rho_0} \frac{\left| \left\langle Q \right\rangle_0 \right|}{\pi R^2}
\label{eqn:V_LCII}
\end{equation}
and $L$ vanishes asymptotically according to Eq. \ref{eqn:L_LCI} with $K = \kappa_{\gamma}$ as $\mathcal{N}_R \rightarrow 0$. At this stage, we could hypothetically extract an estimate of $\eta_S + \eta_R$, but we could not obtain an independent estimate of $\eta_R$ from $\Delta \mathcal{L}_0$ because of the additional dependence of $L$ on $\phi$ in Eq. \ref{eqn:L_LCI}.

\subsubsection{IIciii. Limiting case III: $\kappa_{\phi}/\kappa_{\gamma} \rightarrow 0$}

On the other hand, we may consider the case that $\kappa_{\phi}/\kappa_{\gamma} \rightarrow 0$ while allowing the other dimensionless parameters to remain finite. In this case, we again obtain $K = \kappa_{\gamma}$ from Eq. \ref{eqn:K_general}, as in limiting case II (Eq. \ref{eqn:K_LCII}). We thus have again Eq. \ref{eqn:V_LCII} for $V$ and could in principle extract an estimate of $\eta_S + \eta_R$ from an exponential fit of the velocity profile. In this regime, $L$ is given by
\begin{equation}
L = \frac{2 \mathcal{N}_R}{1 - \mathcal{A}_e} \kappa_{\gamma}^{-1} V
\label{eqn:L_LCIII}
\end{equation}
The dependence of $L$ on $\mathcal{A}_e$ in addition to $\mathcal{N}_R$ indicates that we could not obtain an independent estimate of $\eta_R$ from $\Delta \mathcal{L}_0$ in this regime either.

\subsubsection{IIciv. Limiting case IV: $\mathcal{A}_e, \, \kappa_{\phi}/\kappa_{\gamma} \rightarrow 0$}

Finally, we examine the case that both $\kappa_{\phi} / \kappa_{\gamma}$ and $\mathcal{A}_e$ tend to zero. We again obtain Eq. \ref{eqn:K_LCII}, $K = \kappa_{\gamma}$, and Eq. \ref{eqn:V_LCII}. Additionally, we obtain for $L$
\begin{equation}
L = 2 \mathcal{N}_R \kappa_{\gamma}^{-1} V
\label{eqn:L_LCIV}
\end{equation}
Thus, in this regime, and only this regime, we may obtain an estimate of $\eta_S + \eta_R$ from an exponential fit of the velocity profile and an independent estimate of $\eta_R$ from the spin density profile. Our successful comparison of such estimates to forced runs in main text Fig. 2 indicates that we are indeed in this regime.

From the fact that we are necessarily in this limiting regime, we infer that the boundary current verifies the following balance equations:
\begin{equation}
0 = - \partial_r p_0 + S_{\eta_O} \left| \eta_O \right| \kappa_{\gamma}^2 v_0
\label{eqn:pressure_eqn_LCIV}
\end{equation}
\begin{equation}
\kappa_{\gamma}^2 v_0 = \partial_r^2 v_0
\label{eqn:velocity_eqn_LCIV}
\end{equation}
and
\begin{equation}
\kappa_{\gamma}^2 \Delta \mathcal{L}_0 = 2 \mathcal{N}_R \partial_r v_0
\label{eqn:spin_eqn_LCIV}
\end{equation}
These equations are notable for two reasons. First, we see that the spin is entirely decoupled from the linear momentum equations (Eqs. \ref{eqn:pressure_eqn_LCIV} and \ref{eqn:velocity_eqn_LCIV}), though the vorticity $\omega = \partial_r v_0$ acts as a source in the spin equation (Eq. \ref{eqn:spin_eqn_LCIV}). From this, we conclude that the spin acts as a simple passive tracer, indicating the local value of the vorticity by the relation $\Delta \mathcal{L}_0 = \left( 2 \mathcal{N}_R / \kappa_{\gamma}^2 \right) \omega$. In essence, the spin acts like a passive `dye' that follows the fluid vorticity rather than the velocity itself, allowing us to treat the linear momentum equations entirely separately from the spin dynamics.

\subsection{IId. Analogy to a coastal quasigeostrophic current}

Second, and as discussed in the main text, this problem maps exactly onto a problem in coastal oceanography -- that of a coastally bound quasigeostrophic current, typically formed by the local injection of relatively buoyant water into a coastal environment from, for example, an estuarine source or adjacent oceanographic basin \cite{Alex_ARFM, Alex_JFM, coastal_current1, coastal_current2}. In this context, the relevant rotational Korteweg-de Vries (KdV) equations for a current of thickness $h_0$ are
\begin{equation}
0 = - \partial_r p_0 + \rho_0 S_f \left| f \right| v_0
\label{eqn:KdV_current_umom}
\end{equation}
and
\begin{equation}
\frac{C_{\rm lin} \rho_0 v_0}{h_0} = \eta_S^{\rm eddy} \partial_r^2 v_0
\label{eqn:KdV_current_vmom}
\end{equation}
where $f$ is the local Coriolis frequency of sign $S_f$, $\eta_S^{\rm eddy}$ is an effective `eddy viscosity' parameterizing the turbulent shear stress, and $C_{\rm lin}$ is a linear drag coefficient, characterizing the friction either at the interface between the buoyant current and a denser quiescent fluid below in the case of a surface-trapped current or at the base of the water column in the case of a bottom-trapped current \cite{coastal_current1, coastal_current1}. We see that these equations are of the same form as Eqs. \ref{eqn:pressure_eqn_LCIV} and \ref{eqn:velocity_eqn_LCIV} and become identical under the identifications $f \rightarrow \eta_O \kappa_{\gamma}^2 / \rho_0$, $C_{\rm lin} / h_0 \rightarrow \gamma / m$ and $\eta_S^{\rm eddy} \rightarrow \eta_S + \eta_R $.

Eq. \ref{eqn:KdV_current_umom} represents a so-called {\it quasigeostrophic} balance between a pressure gradient and the Coriolis force (quasi because it holds in only one direction). The geostrophic balance $0 = - \nabla p + \rho f {\boldsymbol u}^*$ is the fundamental unforced steady state of the ocean and atmosphere \cite{Gill}. Much like an odd viscous fluid, the presence of the parity and time-reversal symmetry breaking parameter $f$ leads to the formation of spontaneous unforced currents and a wealth of anomalous transport processes in the atmosphere and ocean \cite{Gill}. Quasigeostrophy, on the other hand, pertains not only to the buoyant coastal currents discussed here but to, for example, the intensified western boundary currents forming in the ocean, the most famous examples of which being the Gulf Stream and Kuroshio Current, forming on the eastern coasts of North America and southeast Asia, respectively \cite{Gill}.

Eqs. \ref{eqn:KdV_current_umom} and \ref{eqn:KdV_current_vmom} are subjected to an integral constraint representing the `conservation of freshwater':
\begin{equation}
\dot{V}_{\rm buoyant} = 2 \pi R \int_0^R {\rm d} r \, \frac{\Delta \rho}{\rho_{\rm dense}} v_0
\label{eqn:Vbuoyant}
\end{equation}
where $\dot{V}_{\rm buoyant}$ is the volumetric flux of buoyant water from the injection source into the boundary current and $\Delta \rho \equiv \rho_{\rm dense} - \rho_0$ is the difference in densities between the buoyant current and the denser water in the receiving basin of density $\rho_{\rm dense}$ \cite{Alex_ARFM, Alex_JFM}. This likewise is analogous to the integral constraint on Eqs. \ref{eqn:pressure_eqn_LCIV} and \ref{eqn:velocity_eqn_LCIV} given in Eq. \ref{eqn:orbital_simple}. The exact structural forms of Eqs. \ref{eqn:orbital_simple} and \ref{eqn:Vbuoyant} are trivially distinct, but the conceptual analogy is exact. In the odd viscous case, the current must form to conserve angular momentum, and in the geophysical case, the current must form to conserve freshwater.

Consider the coastal oceanographic scenario in which an estuarine source discharges freshwater to the coastal ocean through a narrow outlet perpendicular to the local coastline. In this case, the injected momentum will be rapidly dissipated into the receiving quiescent basin and the buoyant plume that forms will be forced by the presence of the Coriolis force to break symmetry over some characteristic length -- as it turns out, this length is the Rossby radius of deformation $\ell_R$, which will be defined and discussed extensively in the next section -- and propagate in a preferred direction along the coast \cite{Alex_ARFM}.

This preferred direction is said to be given by the direction of propagation of coastally bound Kelvin waves, a chiral wave mode obtained from the unsteady counterparts to Eqs. \ref{eqn:KdV_current_umom} and \ref{eqn:KdV_current_vmom} \cite{Gill, science_topology, continuum_topology}. In this case, the chirality of this boundary mode is in accord with the band topographic notion of bulk-boundary correspondence: the presence of the `topological order parameter' $f$ breaks parity and time-reversal symmetry in the bulk of the fluid ({\it i.e.}, in the ocean and atmosphere), introducing a band gap between bulk low frequency Rossby and high frequency Poincar\'{e} waves and therefore a nontrivial band topology \cite{science_topology}. This nontrivial topology can be associated for each band with an integer-valued topological invariant known as the Chern number through a generalized form of the Gauss-Bonnet theorem, and the difference in these numbers across either the topological interface at the equator where $f$ changes sign or at continental boundaries where $f$ effectively vanishes on the continent corresponds to the net number of permitted chiral modes at the boundary or interface. That is, $\Delta N_n = \Delta \mathcal{C}_n$, where $\Delta N_n \equiv N_n^{\rm CCW} - N_n^{\rm CW}$ is the net number of permitted chiral edge modes associated with the $n$-th band, and $\Delta \mathcal{C}_n$ is the change in Chern number $\mathcal{C}_n$ associated with the $n$-th band across the topological interface or boundary. See Refs. \cite{topology_soft_matter, fuchs} for reviews of band topology in active and condensed matter contexts, respectively, and Refs. \cite{boundary_waves, science_topology, continuum_topology} for applications of band topology to geophysical and confined odd viscous flows.

It is this striking analogy between the odd viscous boundary current and the quasigeostrophic buoyant boundary current that leads us to search for Kelvin-like chiral wave modes in the confined odd viscous fluid. The absence of a nonvanishing Coriolis-like parameter in the bulk of the odd viscous fluid results in a closed band gap and hence topologically trivial bulk bands \cite{boundary_waves}, suggesting that a net number of chiral boundary wave modes is prohibited by bulk-boundary correspondence. However, motivated by the observation that the boundary-trapping of the odd viscous current leads to the spontaneous emergence of a Coriolis-like parameter $\eta_O \kappa_{\gamma}^2 / \rho_0$, we suspect that a similar mechanism might occur in boundary-trapped wave modes, leading to the emergence of Kelvin-like chiral waves and an apparent violation of bulk-boundary correspondence. We show in the following section that this is indeed the case.

\section{III. Derivation of odd viscous Kelvin wave mode}
\label{sec:Kelvin}

\subsection{IIIa. Hydrodynamic equations in the weakly compressible limit}

In this section we derive the form of the acoustic wave bands permitted at the boundary of our system. We begin by deriving the weakly compressible hydrodynamic equations. We will consider weak perturbations in density and velocity about the boundary current solution in the limit that $\phi$ is negligible and the spin is therefore decoupled from the mass and linear momentum conservation equations. In this regime, the boundary current satisfies
\begin{equation}
\nabla \cdot {\boldsymbol u}_0 = 0
\label{eqn:boundary_current_mass}
\end{equation}
and
\begin{equation}
\rho_0 {\boldsymbol u}_0 \cdot \nabla {\boldsymbol u}_0 = - \frac{\gamma}{m} \rho_0 {\boldsymbol u}_0 - \nabla p_0 + \left( \eta_S + \eta_R \right) \nabla^2 {\boldsymbol u}_0 + \eta_O \nabla^2 {\boldsymbol u}_0^*
\label{eqn:boundary_current_momentum}
\end{equation}
We now introduce weak perturbations to the density and velocity: $\Delta \rho \equiv \rho - \rho_0$ and $\Delta {\boldsymbol u} \equiv {\boldsymbol u} - {\boldsymbol u}_0$, with $\left| \Delta \rho \right| \ll \rho_0$ and $\left| \Delta {\boldsymbol u} \right| \ll \left| {\boldsymbol u}_0 \right|$. Additionally, we assume that deviations in the pressure from $p_0$ are related to $\Delta \rho$ by $\Delta p \equiv p - p_0 = c^2 \Delta \rho$, where $c^2 \equiv \partial_\rho p |_{\rho_0}$ is the square of the bulk speed of sound in the absence of odd viscosity (or for long wavelengths in the presence of odd viscosity). By inserting these decompositions into Eqs. \ref{eqn:continuity_final} and \ref{eqn:momentum_final}, retaining only terms linear in the perturbations, and taking advantage of Eqs. \ref{eqn:boundary_current_mass} and \ref{eqn:boundary_current_momentum}, we obtain
\begin{equation}
\left( \partial_t + {\boldsymbol u}_0 \cdot \nabla \right) \Delta \rho = - \rho_0 \nabla \cdot \Delta {\boldsymbol u}
\label{eqn:wave_continuity}
\end{equation}
and
\begin{equation}
\begin{split}
\rho_0 \left( \partial_t + {\boldsymbol u}_0 \cdot \nabla \right) &\Delta {\boldsymbol u} + \left( \Delta \rho {\boldsymbol u}_0 + \rho_0 \Delta {\boldsymbol u} \right) \cdot \nabla {\boldsymbol u}_0 \\
= &- \frac{\gamma}{m} \left( \Delta \rho {\boldsymbol u}_0 + \rho_0 \Delta {\boldsymbol u} \right) - c^2 \nabla \Delta \rho \\
&+ \left( \eta_S + \eta_R \right) \nabla^2 \Delta {\boldsymbol u} + \eta_O \nabla^2 \Delta {\boldsymbol u}^* \\
&- \left( \eta_R \nabla + \lambda \nabla^* \right) \nabla \cdot \Delta {\boldsymbol u}
\end{split}
\label{eqn:wave_momentum_frictional}
\end{equation}
As is common in wave analysis, we will assume that frictional terms are negligible. This requires that the linear friction be small and that the odd viscosity be large in magnitude compared to $\eta_S$ and $\eta_R$. Additionally, for simplicity we will assume $\left| \eta_O \right| \gg \lambda$; this assumption only simplifies the calculation and does not affect our conclusions. Previous studies have investigated the influence of frictional terms on wave motions in odd viscous fluids \cite{boundary_waves} and on coastal Kelvin waves \cite{friction_kelvin}, and we anticipate the influence of such terms to modify our results only quantitatively. With these additional assumptions, we obtain for the momentum equation
\begin{equation}
\begin{split}
\rho_0 \left( \partial_t + {\boldsymbol u}_0 \cdot \nabla \right) &\Delta {\boldsymbol u} + \left( \Delta \rho {\boldsymbol u}_0 + \rho_0 \Delta {\boldsymbol u} \right) \cdot \nabla {\boldsymbol u}_0 \\
&= - c^2 \nabla \Delta \rho + \eta_O \nabla^2 \Delta {\boldsymbol u}^*
\end{split}
\label{eqn:wave_momentum}
\end{equation}
Eqs. \ref{eqn:wave_continuity} and \ref{eqn:wave_momentum} are the wave equations that we will analyze in what follows.

\subsection{IIIb. Bulk waves}

The bulk wave modes of an odd viscous fluid have been derived elsewhere \cite{boundary_waves}. In this section, we outline the straightforward approach to derive these modes and list the results. The bulk modes are obtained by taking the Fourier transforms of Eqs. \ref{eqn:wave_continuity} and \ref{eqn:wave_momentum} in the absence of the boundary terms containing ${\boldsymbol u}_0$. Solving the characteristic equation of the eigenvalue problem so obtained gives the following bulk frequencies:
\begin{equation}
\omega_0^{\rm bulk} = 0; \quad \omega_{\pm}^{\rm bulk} = \pm c k \sqrt{ 1 + \left( \frac{k}{\kappa_R} \right)^2 }
\label{eqn:bulk_bands}
\end{equation}
where $\kappa_R \equiv \rho_0 c / \left| \eta_O \right|$ is the inverse of the Rossby radius of deformation. Interestingly, the zero-frequency mode corresponds to a nontrivial steady state motion in which the density anomaly is related to the vorticity via $\Delta \rho = (\eta_O / c^2) \omega$ \cite{boundary_waves}. We note that, in the long wavelength limit $k \rightarrow 0$, the $\omega_{\pm}^{\rm bulk}$ dispersion relations vanish asymptotically as
\begin{equation}
\omega_{\pm}^{\rm bulk} \sim \pm c k
\label{eqn:bulk_bands_long_wavelength}
\end{equation}
That is, we recover the ordinary bulk acoustic dispersion relation in the absence of odd viscosity in the limit that $k \rightarrow 0$.

These bulk bands do not permit a global band gap because they all coincide at $k = 0$, where $\omega_0^{\rm bulk} = \omega_+^{\rm bulk} = \omega_-^{\rm bulk} = 0$. From this, we infer that the band topologies are trivial and a net number of chiral edge states should not be permitted according to bulk-boundary correspondence \cite{topology_soft_matter, boundary_waves, fuchs}. Just beyond the $k = 0$ point, the $\omega_{\pm}^{\rm bulk}$ bands grow linearly (Eq. \ref{eqn:bulk_bands_long_wavelength}), similar in structure to a Dirac point \cite{waves_Kranthi, continuum_topology, gauge_theory}. Main text Fig. 3 shows a plot of the bulk bands along with the chiral Kelvin edge band that we derive in the following section.

\subsection{IIIc. One-dimensional boundary waves}

Interfacially bound wave modes are those whose amplitudes differ appreciably from zero only over some distance $\kappa_r^{-1}$ from the boundary. Such waves can satisfy the no flux condition at the boundary only if the amplitude of the radial component $\Delta u$ of their fluctuating velocity vanishes. In this regime, Eqs. \ref{eqn:wave_continuity} and \ref{eqn:wave_momentum} in cylindrical coordinates reduce to
\begin{equation}
\left( \partial_t + v_0 \partial_y \right) \Delta \rho = - \rho_0 \partial_y \Delta v
\label{eqn:continuity_bound}
\end{equation}
\begin{equation}
\begin{split}
- \frac{v_0}{r} &\left( 2 \rho_0 \Delta v + v_0 \Delta \rho \right) = - c^2 \partial_r \Delta \rho \\
&\quad \quad \quad+ S_{\eta_O} \left| \eta_O \right| \left( \partial_r^2 + \partial_y^2 + \frac{\partial_r}{r} - \frac{1}{r^2} \right) \Delta v
\end{split}
\label{eqn:umomentum_bound}
\end{equation}
and
\begin{equation}
\rho_0 \left( \partial_t + v_0 \partial_y \right) \Delta v = - c^2 \partial_y \Delta \rho + S_{\eta_O} \left| \eta_O \right| 2 \frac{\partial_y \Delta v}{r}
\label{eqn:vmomentum_bound}
\end{equation}
where $\Delta v$ is the tangential ($\theta$) component of the fluctuating velocity, $v_0$ is the tangential (and only nonvanishing) component of the boundary current velocity, and we have defined the circumferential coordinate $y \equiv r \theta$, noting that $\partial_{\theta} |_{r,t} \equiv r \partial_y |_{r,t}$. With this notation, we must be careful to remember that $\partial_y$ and $\partial_r$ do not commute. These equations represent, respectively, a statement of mass continuity, the radial balance between centrifugal force and the radial gradients of the acoustic and odd viscous pressures, and the tangential momentum balance. In the latter equation, the final term represents an odd viscous pressure gradient along the direction of flow that is induced by the radius of curvature of a particular circumferential wave characteristic. We search for solutions to these equations in a circular confinement of radius $R$.

We begin by deriving the wave equations satisfied by the anomalies in density and tangential velocity. To do so, we first note that Eqs. \ref{eqn:continuity_bound} and \ref{eqn:vmomentum_bound} contain derivatives only in $t$ and $y$; hence, we may use these equations to solve for the periodic motions of $\Delta \rho$ and $\Delta v$ fully decoupled from radial gradients. We first take the $t$ and $y$ derivatives of these equations to obtain
\begin{equation}
\left( \partial_t^2 + v_0 \partial_t \partial_y \right) \Delta \rho = - \rho_0 \partial_t \partial_y \Delta v
\label{eqn:dt_continuity}
\end{equation}
\begin{equation}
\left( \partial_t \partial_y + v_0 \partial_y^2 \right) \Delta \rho = - \rho_0 \partial_y^2 \Delta v
\label{eqn:dy_continuity}
\end{equation}
\begin{equation}
\rho_0 \left( \partial_t^2 + v_0 \partial_t \partial_y \right) \Delta v = - c^2 \partial_t \partial_y \Delta \rho + S_{\eta_O} 2 \left| \eta_O \right| \frac{\partial_t \partial_y \Delta v}{r}
\label{eqn:dt_vmomentum}
\end{equation}
and
\begin{equation}
\rho_0 \left( \partial_t \partial_y + v_0 \partial_y^2 \right) \Delta v = - c^2 \partial_y^2 \Delta \rho + S_{\eta_O} 2 \left| \eta_O \right| \frac{\partial_y^2 \Delta v}{r}
\label{eqn:dy_vmomentum}
\end{equation}
In order to obtain the wave equation for $\Delta \rho$, we solve Eq. \ref{eqn:dy_vmomentum} for $\partial_t \partial_y \Delta v$, insert the result into Eq. \ref{eqn:dt_continuity}, and eliminate the remaining factors of $\partial_y^2 \Delta v$ using Eq. \ref{eqn:dy_continuity}. Similarly, to isolate $\Delta v$, we solve Eq. \ref{eqn:dy_continuity} for $\partial_t \partial_y \Delta \rho$, insert the result into Eq. \ref{eqn:dt_vmomentum}, and eliminate the remaining factor of $\partial_y^2 \Delta \rho$ using Eq. \ref{eqn:dy_vmomentum}. In both cases, we obtain a wave equation of the form
\begin{equation}
\begin{split}
&\left( \partial_t^2 + S_{\eta_O} c \, {\rm Ma} \frac{v_0}{V} \partial_t \partial_y \right) \psi = c^2 \partial_y^2 \psi \\
&+ S_{\eta_O} 2 c \left[ \left( \kappa_R r \right)^{-1} - \frac{1}{2} {\rm Ma} \frac{v_0}{V} \right] \\
& \quad \quad \quad \quad \times \left( \partial_t \partial_y + S_{\eta_O} c \, {\rm Ma} \frac{v_0}{V} \partial_y^2 \right) \psi
\end{split}
\label{eqn:wave_equation}
\end{equation}
where we have introduced the inverse Rossby radius of deformation $\kappa_R \equiv \rho_0 c / | \eta_O |$. The term $S_{\eta_O} c \, {\rm Ma} \, v_0 / V = v_0$ is really just the background boundary current flow $v_0$; we have written it in this way to make explicit its dependence on the sign of the odd viscosity and the Mach number, defined as
\begin{equation}
{\rm Ma} \equiv \frac{\left| V \right|}{c} = \frac{\kappa_{\gamma}}{2 \rho_0 c} \frac{\left| \left\langle Q \right\rangle_0 \right|}{\pi R^2}
\label{eqn:Ma}
\end{equation}
where we have made use of Eqs. \ref{eqn:friction_lengths} and \ref{eqn:V_LCII}. The ratio $v_0/V$ is independent of the sign of the odd viscosity and given by, from Eqs. \ref{eqn:exponential_form_boundary_current} and \ref{eqn:K_LCII},
\begin{equation}
\frac{v_0}{V} = e^{- \kappa_{\gamma} \left( R - r \right)} \equiv e^{- \frac{\kappa_{\gamma}}{\kappa_R} \kappa_R \left( R - r \right)}
\label{eqn:v_ratio}
\end{equation}

Because only derivatives of $t$ and $y$ appear in Eq. \ref{eqn:wave_equation}, we may solve it as a second order PDE with constant coefficients; however, the coefficients will in reality depend on the local value of $r$ -- both through the terms proportional to $v_0(r)/V$, induced by coupling to the boundary current flow, and the term proportional to $\left( \kappa_R r \right)^{-1}$, induced by the curvature-dependent odd viscous pressure gradient appearing in Eq. \ref{eqn:vmomentum_bound}.

Eq. \ref{eqn:wave_equation} is subject to the periodicity condition $\psi \left( y + 2 n \pi r \right) = \psi \left( y \right)$ for integers $n$. Solutions are therefore of the form
\begin{equation}
\psi = e^{i \left( k_y^n (r) y - \omega_n (r) t \right)}; \quad k_y^n (r) = \frac{n}{r}
\label{eqn:psi}
\end{equation}
The dependence of $k_y^n$ on $r$ is dictated by the periodicity condition, while the dependence of $\omega_n / c k_y^n$ on $r$ is determined by the $r$-dependent coefficients in Eq. \ref{eqn:wave_equation}. In particular, inserting Eq. \ref{eqn:psi} into Eq. \ref{eqn:wave_equation} gives the following dispersion relation:
\begin{equation}
\begin{split}
\frac{\omega_n (r)}{c k_y^n (r)} &= S_{\eta_O} S_{\rm band} \sqrt{1 + \left( \kappa_R r \right)^{-2}} \\
&\quad \quad \quad- S_{\eta_O} \left[ \left( \kappa_R r \right)^{-1} - {\rm Ma} e^{- \frac{\kappa_{\gamma}}{\kappa_R} \kappa_R \left( R - r \right)} \right]
\end{split}
\label{eqn:omega_cky}
\end{equation}
where $S_{\rm band} \equiv S_{\omega} S_{\eta_O}$, and $S_{\omega}$ denotes the sign of the branch of the dispersion relation. The character of the slope function depends on the relative sign of the odd viscosity and frequency band through the prefactor $S_{\rm band}$. In what follows, we will refer to the $S_{\rm band} = +1$ solution as the pchiral band and the $S_{\rm band} = -1$ solution as the nchiral band. 

Eq. \ref{eqn:omega_cky} indicates that, in the limit ${\rm Ma} \ll 1$, we recover a one-dimensional form of the ordinary bulk acoustic dispersion relation $\omega = S_{\omega} c k$, except that the slope of this relation is modified by a local function of the radial position; this is entirely a consequence of the odd viscous pressure.

In what follows, the dispersion relation at the boundary will be important. Defining $\omega_n^R \equiv \omega_n \left( \kappa_R R \right)$, we have from Eq. \ref{eqn:omega_cky}
\begin{equation}
\begin{split}
\frac{\omega_n^R}{c k_y^n(R)} &= S_{\eta_O} S_{\rm band} \sqrt{1 + \left( \kappa_R R \right)^{-2}} \\
&\quad \quad \quad \quad- S_{\eta_O} \left[ \left( \kappa_R R \right)^{-1} - {\rm Ma} \right] \\
&\xrightarrow[\kappa_R R \rightarrow \infty]{} S_{\eta_O} \left( S_{\rm band} + {\rm Ma} \right)
\end{split}
\label{eqn:boundary_dispersion}
\end{equation}
from which we see that we obtain the one-dimensional, Doppler-shifted bulk dispersion relation at the boundary asymptotically as the system size diverges. Interestingly, we also note that, if the Mach number defined in Eq. \ref{eqn:Ma} is greater than unity, a supersonic regime obtains and both chiral bands propagate in the same direction determined by the sign of the odd viscosity with the dispersions $\omega_n^R / c k_y^n(R) = S_{\eta_O} \left( {\rm Ma} \pm 1 \right)$. The value of the Mach number is ultimately determined by the imposed density and activity (Eq. \ref{eqn:Ma}). Whether and under what conditions supersonic flow can occur is beyond the scope of this study.

\begin{figure}[h]
	\centerline{\includegraphics[width=8.5cm]{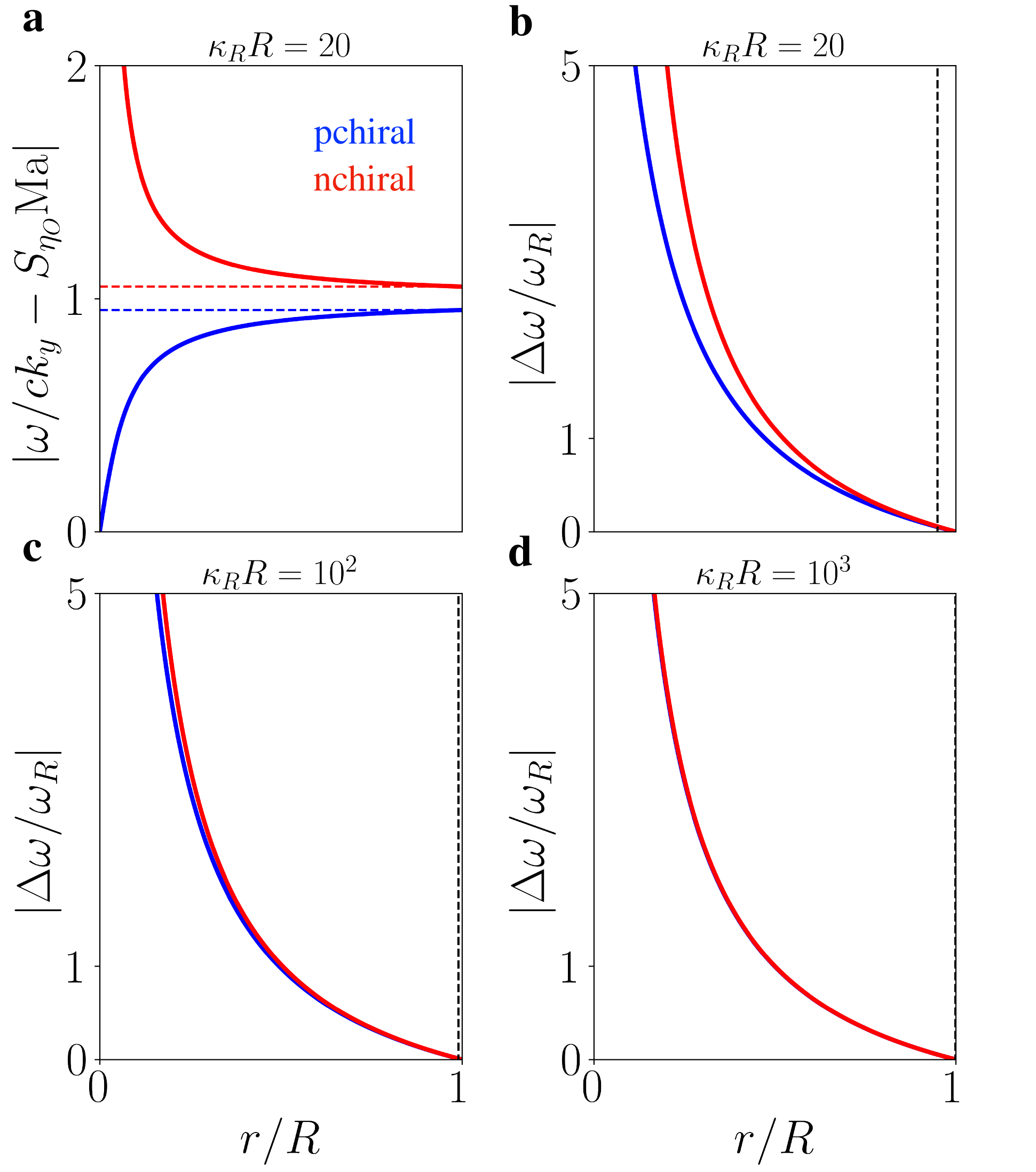}}
    \caption{Radial dependence of the frequency. {\bf a}, Slope of the dispersion relation as a function of radius for pchiral and nchiral bands. {\bf b}, Deviation of frequency from its value at the boundary as a fraction of the boundary value and a function of radial position. {\bf c}, {\bf d}, Same as in {\bf b}, but for system sizes $\kappa_R R = 10^2$ and $10^3$, respectively. The dashed horizontal lines in {\bf a} indicate the limiting value of the slope for $\kappa_R R = 20$, and the dashed vertical lines in {\bf b} through {\bf d} indicate the position of the Rossby radius. As in {\bf a}, the Doppler-shift contribution to the frequencies in {\bf b} through {\bf d} have been subtracted out.}
\label{fig:Fig2}
\end{figure}

For finite system sizes the symmetry between the nchiral and pchiral bands is broken -- that is, $|\omega_{\rm pband}^R / c k_y^n (R) - S_{\eta_O} {\rm Ma} | \neq |\omega_{\rm nband}^R / c k_y^n (R) - S_{\eta_O} {\rm Ma} |$. This is a consequence of the dependence of the odd viscous pressure gradient appearing in Eq. \ref{eqn:vmomentum_bound} on the sign of $\eta_O$. These results are illustrated in Figs. \ref{fig:Fig2}a and b for a system size $\kappa_R R = 20$, as in the main text, and for larger system sizes in Figs. \ref{fig:Fig2}c and d.

The fact that both $\Delta \rho$ and $v$ satisfy Eq. \ref{eqn:wave_equation} reveals that solutions must be of the form
\begin{equation}
\begin{Bmatrix}
\Delta \rho \\
\Delta v
\end{Bmatrix}
=
\begin{Bmatrix}
\phi_{\Delta \rho} (r) \\
\phi_{\Delta v} (r)
\end{Bmatrix}
\psi \left( \theta, t; n, \omega_n (r) \right) +
\begin{Bmatrix}
b_{\Delta \rho} (r) \\
b_{\Delta v} (r)
\end{Bmatrix}
\label{eqn:solution_form1}
\end{equation}
That is, $\psi$ may be attenuated and shifted by functions of $r$, and these functions may in principle be distinct for $\Delta \rho$ and $u$. However, continuity, Eq. \ref{eqn:continuity_bound}, requires that $\phi_{\Delta \rho}(r)$ and $\phi_{\Delta v} (r)$ are related by
\begin{equation}
\begin{split}
\frac{\phi_{\Delta \rho}(r) / \rho_0}{\phi_{\Delta v}(r) / c} &= \left[ \frac{\omega_n(r)}{c k_y^n(r)} - S_{\eta_O} {\rm Ma} \frac{v_0}{V} \right]^{-1} \\
&= S_{\eta_O} \left[ S_{\rm band} \sqrt{1 + \left( \kappa_R r \right)^{-2}} - \left( \kappa_R r \right)^{-1} \right]^{-1}
\end{split}
\label{eqn:attenuation_relation}
\end{equation}
where we have inserted the dispersion relation, Eq. \ref{eqn:omega_cky}, in the second equality. This leaves a single equation, Eq. \ref{eqn:umomentum_bound}, to specify the three remaining radial functions $\phi_{\Delta v} (r) \equiv \hat{v} c \phi (r)$, $b_{\Delta \rho} (r)$, and $b_{\Delta v} (r)$, indicating that only one of them, the amplitude attenuation function $\phi(r)$, may be nonzero. Thus, our solutions must be of the form
\begin{equation}
\begin{Bmatrix}
\Delta \rho \\
\Delta v
\end{Bmatrix}
=\hat{v}
\begin{Bmatrix}
c \\
S_{\eta_O} S_{\rm band} \Phi(\kappa_R r) \rho_0
\end{Bmatrix}
\phi(r) \psi\left( \theta , t; n, \omega_n \left( r \right) \right)
\label{eqn:solution_form2}
\end{equation}
with
\begin{equation}
\Phi(\kappa_R r) = \left( \sqrt{1 + \left( \kappa_R r \right)^{-2}} - S_{\rm band} \left( \kappa_R r \right)^{-1} \right)^{-1} > 1
\label{eqn:hat_relation}
\end{equation}
a dimensionless function that is greater than one for all $\kappa_R r$ and approaches unity as $\kappa_R r \rightarrow \infty$ for $S_{\rm band} = \pm 1$.

Inserting Eqs. \ref{eqn:solution_form2} and \ref{eqn:hat_relation} into Eq. \ref{eqn:umomentum_bound}, we finally obtain the equation governing the radial structure of our bound wave modes:
\begin{equation}
\kappa_R^{-2} \partial_r^2 \left( \phi \psi \right) - S_{\rm band} a_1(\kappa_R r) \kappa_R^{-1} \partial_r \left( \phi \psi \right) - a_0^n(\kappa_R r) \phi \psi = 0
\label{eqn:radial_structure}
\end{equation}
with the radially dependent dimensionless coefficients $a_1(\kappa_R r)$ and $a_0^n(\kappa_R r)$ given by
\begin{equation}
a_1(\kappa_R r) \equiv \Phi(\kappa_R r) - S_{\rm band} \left( \kappa_R r \right)^{-1} > 1
\label{eqn:b_defn}
\end{equation}
and
\begin{equation}
\begin{split}
a_0^n \left( \kappa_R r \right) \equiv &\frac{n^2 + 1}{\left( \kappa_R r \right)^2} + S_{\rm band} \Phi' \left( \kappa_R r \right) \\
&\quad \quad \quad- \frac{{\rm Ma}}{\kappa_R r} \frac{v_0}{V} \left[ 2 + S_{\rm band} {\rm Ma} \frac{v_0}{V} \Phi \left( \kappa_R r \right) \right]
\end{split}
\label{eqn:c_defn}
\end{equation}
where $\Phi'\left( \kappa_R r \right) \equiv d\Phi/d\left(\kappa_R r \right)$. Note that $a_1 \left( \kappa_R r \right)$ is also greater than one for all finite $\kappa_R r$, tending to unity as $\kappa_R r \rightarrow \infty$. Because of the radial dependence of the frequency, we cannot immediately factor out the phase $\psi$. This equation will in general have two solutions, one decaying from the boundary and one decaying from the origin; the latter solution is unphysical and will be discarded. We may integrate this equation for $\phi \psi$ starting from $\kappa_R r = \kappa_R R$ with the boundary condition $\phi \psi |_{\kappa_R R} = {\rm exp} \left[ i \left( n \theta - \omega_n (r) t \right) \right]$. The result for $\phi(r)$ must of course be dependent only on $r$ and therefore independent of the phase we choose at the boundary. We may therefore take $\theta = 0$ and $t = 0$ at the boundary to obtain a solution that depends only on $r$, satisfying
\begin{equation}
\kappa_R^{-2} \partial_r^2 \phi - S_{\rm band} a_1\left( \kappa_R r \right) \kappa_R^{-1} \partial_r \phi - a_0^n\left( \kappa_R r \right) \phi = 0
\label{eqn:radial_structure_nophase}
\end{equation}
with $\phi \left( \kappa_R R \right) = 1$. However, this is the equation that would be obtained if $\psi$ were independent of $r$ and could therefore be factored out of Eq. \ref{eqn:radial_structure}. This will not be true in general for $t \neq 0$. More explicitly, the radial dependence of $\omega$ induces additional radial gradients in Eq. \ref{eqn:radial_structure} that will depend on $t$, leading to a time-dependent solution for $\phi$. This of course would lead to an unacceptable solution of the wave equation, Eq. \ref{eqn:wave_equation}. Furthermore, the additional terms induced by the radial dependence of $\omega$ grow in time. For example, the first order term in Eq. \ref{eqn:radial_structure} will contain a contribution $\phi \partial_r \psi = - i \psi \partial_r \omega_n t \phi$ which oscillates in sign and grows linearly in magnitude with time. This reveals that a slowly varying approximation which assumes that terms proportional to radial gradients of the frequency are negligible is not sufficient to obtain a valid solution to Eq. \ref{eqn:radial_structure}. That is, so long as there exists an appreciable change in $\omega$ across the bound wave, the time-dependent terms in Eq. \ref{eqn:radial_structure} will eventually become non-negligible, indicating a breakdown of the assumption of one-dimensionality. Therefore, we conclude that a one-dimensional interfacially bound wave solution exists only when $\left| \Delta \omega_n \left( \kappa_R r \right) \right| \equiv \left| \omega_n \left( \kappa_R r \right) - \omega_n^R \right| \ll \left| \omega_n^R \right|$. Eqs. \ref{eqn:omega_cky} and \ref{eqn:boundary_dispersion} and Fig. \ref{fig:Fig2} indicate that this is true only if the scale $\kappa_r^{-1}$ of the penetration of the boundary mode into the bulk is small compared to the confinement radius: $\kappa_r R \ll 1$. Physically, we learn that wave modes whose radial extent approaches the radius of curvature must necessarily devolve into two-dimensional bulk waves due to the curvature-induced odd viscous pressure gradient and the radial dependence of $k_y^n = n/r$.

\subsection{IIId. Explicit solution: $\kappa_r R \ll 1$}

We now solve Eqs. \ref{eqn:wave_equation} and \ref{eqn:radial_structure} to second order in $1/\kappa_R R$ in the regime $\kappa_r R \ll 1$. We stress that this is not an approximation to a more general solution; rather, this is the only regime in which solutions to these equations exist and therefore the only regime in which one-dimensional edge modes may exist. The condition $\kappa_r R \ll 1$ allows us to approximate the radius of curvature over the wave as approximately constant and equal to $R$. We obtain from Eqs. \ref{eqn:psi} and \ref{eqn:radial_structure}
\begin{equation}
\psi = e^{i \left( n \theta - \omega_n^R t \right)}
\label{eqn:psi_thin}
\end{equation}
and
\begin{equation}
\kappa_R^{-2} \partial_r^2 \phi - S_{\rm band} a_1^R \kappa_R^{-1} \partial_r \phi - a_0^{n,R} \left( \kappa_R r \right) \phi = 0
\label{eqn:radial_structure_nophase2}
\end{equation}
with $a_1^R \equiv a_1 \left( \kappa_R R \right)$, and
\begin{equation}
\begin{split}
&a_0^{n,R} \left( \kappa_R r \right) \equiv \frac{n^2 + 1}{\left( \kappa_R R \right)^2} + S_{\rm band} \Phi' \left( \kappa_R R \right) \\
&- \frac{{\rm Ma}}{\kappa_R R} e^{- \frac{\kappa_{\gamma}}{\kappa_R} \left( \kappa_R R - \kappa_R r \right)} \left[ 2 + S_{\rm band} {\rm Ma} e^{- 2 \frac{\kappa_{\gamma}}{\kappa_R} \left( \kappa_R R - \kappa_R r \right)} \right]
\end{split}
\label{eqn:bn_defn}
\end{equation}
from Eqs. \ref{eqn:v_ratio} and \ref{eqn:c_defn}.

While we are justified in taking the radius of curvature to be constant in this regime, we cannot neglect the radial dependence of the boundary current {\it a priori}. Though an analytical solution to Eqs. \ref{eqn:radial_structure_nophase2} and \ref{eqn:bn_defn} may be found for a finite Mach number, we will take the limit that ${\rm Ma} \ll 1$ in what follows. This greatly simplifies the analysis and amounts to assuming that the boundary flow is slow enough compared to the bulk speed of sound in this system that it will not couple significantly to the wave modes. This is likely a very good approximation and will not modify our qualitative conclusions. We further retain only the lowest (second) order contribution to $\Phi'\left( \kappa_R R \right)$ in Eq. \ref{eqn:bn_defn}. From Eq. \ref{eqn:hat_relation}, one can show that this is given by $\Phi'\left( \kappa_R R \right) \sim - S_{\rm band} / \left( \kappa_R R \right)^2$. With these approximations, Eq. \ref{eqn:bn_defn} reduces to
\begin{equation}
a_0^{n,R} = \left( \frac{n}{\kappa_R R} \right)^2 \equiv \left( \frac{k_y^n}{\kappa_R} \right)^2
\label{eqn:bn_defn2}
\end{equation}

\begin{figure}[h]
	\centerline{\includegraphics[width=8.5cm]{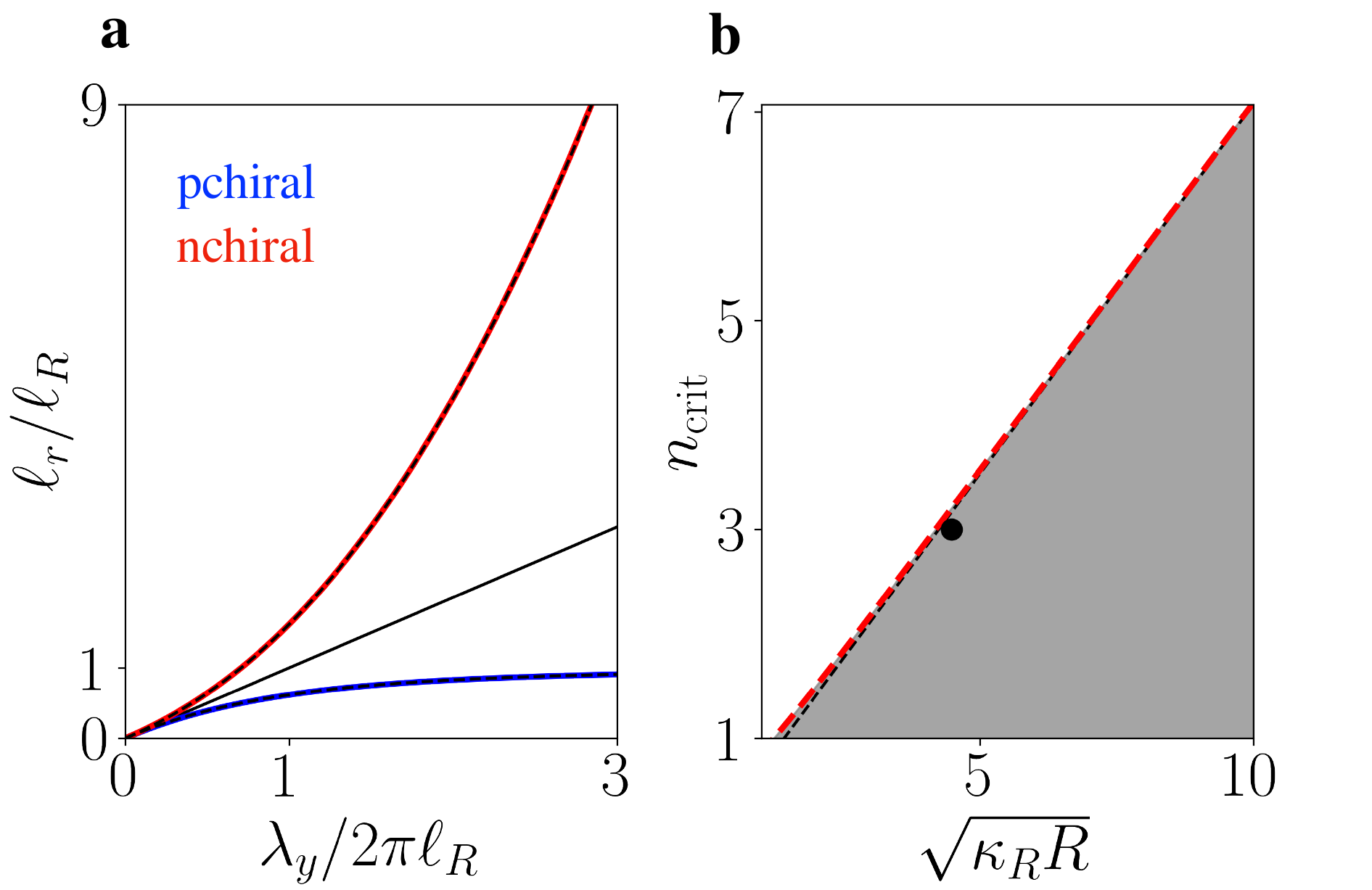}}
    \caption{{\bf a}, Bulk penetration scale as a function of lateral wavelength for $\kappa_R R = 20$ (solid, colored lines) and $\kappa_R R \rightarrow \infty$ (dashed black line). The solid black line indicates a one-to-one relationship. {\bf b}, Requiring $\kappa_r^{\rm nchiral} R \ll 1$ introduces a critical integer wavenumber, plotted here as a function of system size (dashed red). The dashed black line indicates the asymptotic scaling $n_{\rm crit} \sim \sqrt{2 \kappa_R R}/2$, the gray shaded area indicates the region where $\kappa_r^{\rm nchiral} R > 1$, and the dot indicates $\left(\kappa_R R, n \right) = \left( 20, 3 \right)$, corresponding to the pchiral wave solution plotted in the main text.}
\label{fig:Fig3}
\end{figure}

The solution of Eqs. \ref{eqn:radial_structure_nophase2} and \ref{eqn:bn_defn2} is given by
\begin{equation}
\phi(r) = e^{\kappa_r r}
\label{eqn:phi_solution}
\end{equation}
with
\begin{equation}
\frac{\kappa_r}{\kappa_R} = \frac{1}{2} \left[ S_{\rm band} a_1^R + \sqrt{\left( a_1^R \right)^2 + \left( \frac{2 n}{\kappa_R R} \right)^2} \right]
\label{eqn:kappar_solution}
\end{equation}
Equivalently, we can solve for the penetration length $\ell_r \equiv \kappa_r^{-1}$ normalized by the Rossby radius $\ell_R \equiv \kappa_R^{-1}$ to obtain
\begin{equation}
\begin{split}
\left( \frac{n}{\kappa_R R} \right)^2 \frac{\ell_r}{\ell_R} = \frac{1}{2} \left[ - S_{\rm band} a_1^R + \sqrt{\left( a_1^R \right)^2 + \left( \frac{2 n}{\kappa_R R} \right)^2} \right]
\end{split}
\label{eqn:ellr_solution}
\end{equation}
These results are dramatically simplified if we take $\kappa_R R \rightarrow \infty$ but allow for $n / \kappa_R R \equiv k_y / \kappa_R$ to remain finite -- {\it i.e.}, if we allow for waves much shorter than the circumference of the confinement and comparable in wavelength to the Rossby radius. We find for $\kappa_r$ and $\ell_r$ in this regime
\begin{equation}
\frac{\kappa_r}{\kappa_R} = \frac{1}{2} \left[ S_{\rm band} + \sqrt{1 + \left( \frac{2 k_y}{\kappa_R} \right)^2} \right]
\label{eqn:kappar_infinite_system}
\end{equation}
and
\begin{equation}
\frac{\ell_r}{\ell_R} = \frac{1}{2} \left( \frac{\lambda_y}{2 \pi \ell_R} \right)^2 \left[ - S_{\rm band} + \sqrt{1 + \left( \frac{4 \pi \ell_R}{\lambda_y} \right)^2} \right]
\label{eqn:ellr_infinite_system}
\end{equation}
where $\lambda_y \equiv 2 \pi / k_y$ is the tangential wavelength. It turns out that for even moderate system sizes $\kappa_R R \gtrsim 10$, the predictions of Eqs. \ref{eqn:kappar_infinite_system} and \ref{eqn:ellr_infinite_system} do not differ appreciably from those of Eqs. \ref{eqn:kappar_solution} and \ref{eqn:ellr_solution}. This is illustrated in Fig. \ref{fig:Fig3}a, where we plot $\ell_r$ against the tangential wavelength $\lambda_y / 2 \pi$ in units of the Rossby radius for both $\kappa_R R = 20$, the system size investigated in the main text, and $\kappa_R R \rightarrow \infty$.

From Eq. \ref{eqn:ellr_infinite_system} and Fig. \ref{fig:Fig3}a, we learn that for small wavelengths $\lambda_y/2\pi \ll \ell_R$, $\ell_r$ grows linearly with $\lambda_y / 2 \pi$ for both bands. On the other hand, as $\lambda_y/2\pi$ becomes larger than $\ell_R$, $\ell_r^{\rm pband} \rightarrow \ell_R$, while $\ell_r^{\rm nband}$ diverges as $(\lambda_y/2\pi)^2$. Thus, as $\lambda_y$ is increased, the assumption $\kappa_r R \ll 1$ must break down. However, as discussed above, solutions to Eq. \ref{eqn:radial_structure} can exist only in the regime $\kappa_r R \ll 1$, for which $\omega_n$ is approximately constant. Therefore, {\it the divergence of} $\ell_r^{\rm nband}$ {\it with} $\lambda_y^2$ {\it indicates that the nband wave can only exist up to some finite, system-size dependent value of} $\lambda_y$. Nband waves with $\lambda_y$ greater than this critical value {\it must necessarily devolve into two-dimensional bulk waves}.

We can formulate a bound on the critical value of $\lambda_y$ -- or equivalently the critical value of the integer wavenumber $n$ -- by setting $\ell_r^{\rm nband} = R$. In reality, the nband wave must break down long before its thickness approaches $R$, and this will be only a weak lower bound on the true critical value of the integer wavenumber. From either of Eqs. \ref{eqn:kappar_infinite_system} or \ref{eqn:ellr_infinite_system}, we find that the condition for nband waves to exist is
\begin{equation}
n \gg n_{\rm crit} \equiv \frac{1}{2} \sqrt{1 + 2 \kappa_R R } \xrightarrow{\kappa_R R \rightarrow \infty} \infty
\label{eqn:ncrit}
\end{equation}
That is, the critical value of $n$ diverges with the square root of $\kappa_R R$, such that in the limit $\kappa_R R \rightarrow \infty$ all nband waves are entirely eliminated. $n_{\rm crit}$ is plotted in Fig. \ref{fig:Fig3}b. From this expression, one can show that $\lambda_y^{\rm crit} \sim 2 \pi \sqrt{2 \ell_R R}$ as $\kappa_R R \rightarrow \infty$. {\it Thus, we find that for wavelengths approaching the length scale} $2 \pi \sqrt{2 \ell_R R}${\it, nband waves must devolve into two-dimensional bulk waves, and that in the limit of infinite system size, there is no value of the integer wave number} $n$ {\it large enough to allow for nband waves.}

\subsection{IIIe. Comparison to coastal Kelvin wave solution}

Before leaving this section, it is instructive to compare our problem to the oceanographic problem leading to the inherently chiral coastal Kelvin wave solution in order to better clarify the physics that have allowed a net chirality to persist in our system at long wavelengths even in the absence of a bulk band gap. Kelvin waves are solutions to the rotating, frictionless, linearized KdV equations in the presence of a solid boundary \cite{Gill}. They are formulated in terms of a sea surface height anomaly $\Delta h$, measured relative to the mean depth $h_0$, and an alongshore velocity $v$. We consider an exponentially bound Fourier component of the form $\Delta h, v \sim {\rm exp} (\kappa_r r) {\rm exp} (i(k_y y - \omega t))$, where $y$ is the alongshore coordinate. The linearized KdV equations then reduce to
\begin{equation}
- i \omega \hat{\Delta h} = - i h_0 k_y \hat{v}
\label{eqn:KdV_mass}
\end{equation}
\begin{equation}
0 = - \frac{c^2}{h_0} \kappa_r \hat{\Delta h} + S_f \left| f \right| \hat{v}
\label{eqn:KdV_umomentum}
\end{equation}
and
\begin{equation}
- i \omega \hat{v} = - i \frac{c^2}{h_0} k_y \hat{\Delta h}
\label{eqn:KdV_vmomentum}
\end{equation}
where $f$ is the local Coriolis frequency of sign $S_f$ and $c^2 \equiv g h_0$ is the square of the Kelvin wave speed. In this context, the Rossby radius of deformation is defined as $\ell_R \equiv c / |f|$. Eqs. \ref{eqn:KdV_mass} and \ref{eqn:KdV_vmomentum} may be solved to give the dispersion relation $\omega = \pm c k_y$ identical to our dispersion relation in the limit $\kappa_R R \rightarrow \infty$, and Eqs, \ref{eqn:KdV_mass} and \ref{eqn:KdV_umomentum} may be solved to give $\kappa_r = \pm S_f |f|/c \equiv \kappa_R$. Hence, we must choose the root of sign $S_f$ so that the bound waves decay away from the solid boundary. This is the origin of net chirality in the Kelvin wave problem.

Eqs. \ref{eqn:continuity_bound} through \ref{eqn:vmomentum_bound} may likewise be written (schematically) for a bound Fourier component as
\begin{equation}
- i \omega \hat{\Delta \rho} = - i \rho_0 k_y \hat{v}
\label{eqn:continuity_Kelvin_comparison}
\end{equation}
\begin{equation}
0 = - \frac{c^2}{\rho_0} \kappa_r \hat{\Delta \rho} + S_{\eta_O} \frac{c}{\kappa_R} \left( \kappa_r^2 - k_y^2 + \frac{\kappa_r}{r} - \frac{1}{r^2} \right) \hat{v}
\label{eqn:umomentum_Kelvin_comparison}
\end{equation}
and
\begin{equation}
- i \omega \hat{v} = - i \frac{c^2}{\rho_0} k_y \hat{\Delta \rho} + S_{\eta_O} 2 i \frac{c}{\kappa_R} \frac{k_y}{r} \hat{v}
\label{eqn:vmomentum_Kelvin_comparison}
\end{equation}
These equations are strikingly similar to Eqs. \ref{eqn:KdV_mass} through \ref{eqn:KdV_vmomentum} under the identification $\Delta \rho \rightarrow \Delta h$. In particular, the constant term $c \kappa_r^2 / \kappa_R \equiv \left| \eta_O \right| \kappa_r^2 / \rho_0$ looks like a Coriolis term, leading to the identification $\eta_O \kappa_r^2 / \rho_0 \rightarrow f$, similar to that made for the boundary current solution with $\kappa_r \sim \kappa_R$ in place of $\kappa_{\gamma}$. It is the fact that this Coriolis-like term is present in the bound wave and does not vanish as $k_y \rightarrow 0$ that leads to the presence of a Kelvin-like wave mode. Indeed, in the limit that the radius of curvature $\kappa_r R \rightarrow \infty$ and $k_y/\kappa_r \rightarrow 0$, Eqs. \ref{eqn:continuity_Kelvin_comparison} through \ref{eqn:vmomentum_Kelvin_comparison} reduce exactly to Eqs. \ref{eqn:KdV_mass} through \ref{eqn:KdV_vmomentum} under the above identifications for $\Delta h$ and $f$.

The three differences between these sets of equations are the terms proportional to $1/r$ and $1/r^2$ appearing in Eq. \ref{eqn:umomentum_Kelvin_comparison}, the term proportional to $k_y^2$ appearing in the same equation, and the previously discussed odd viscous pressure gradient appearing in Eq. \ref{eqn:vmomentum_Kelvin_comparison} and induced by the radius of curvature of the local wave characteristic. The first difference is trivial; if these terms are non-negligible, they simply modify the structure of the radial confinement from a simple exponential to a more complicated function that still decays as $r \rightarrow 0$. The latter two differences are crucial: Solving Eqs. \ref{eqn:continuity_Kelvin_comparison} through \ref{eqn:vmomentum_Kelvin_comparison} in the case that the radius of curvature diverges but $k_y$ remains finite reveals that it is the presence of the $-k_y^2$ contribution to the Coriolis-like term that allows for a second wave with diverging bulk penetration scale to form, and which therefore breaks the net chirality. On the other hand, it is the curvature-induced odd viscous pressure gradient, along with the periodicity condition $k_y^n = n/r$, that shifts the wave frequency as a function of $r$ and forces these waves to devolve into two-dimensional bulk waves, thus restoring chirality.

We may summarize our results for edge modes in an odd viscous chiral active fluid qualitatively as follows: at small wavelengths, two boundary-trapped modes are present, leading to zero net edge mode chirality. These correspond to the two edge modes observed in an odd viscous fluid in the absence of external rotation presented in the Supplemental Material of Ref. \cite{boundary_waves}. These waves initially grow in extent linearly with the transverse wavelength $\lambda_y$. However, as $\lambda_y$ is increased to a value $\mathcal{O} (\ell_R)$, the lateral extents begin to diverge, with the bulk penetration depth of the pband wave ($\omega \sim S_{\eta_O} c k_y$) saturating at $\ell_R$ and that of the nbound wave ($\omega \sim - S_{\eta_O} c k_y$) growing as $\lambda_y^2$. As $\lambda_y$ approaches a critical value that scales as $2 \pi \sqrt{2 \ell_R R}$, the nband wave begins to feel the effect of the changing local radius of curvature and the curvature-induced odd viscous pressure and devolves into two-dimensional bulk waves, thereby leaving only the pband Kelvin-like mode and restoring net chirality of the edge modes in the absence of a bulk band gap.

\section{IV. Hydrodynamic solution for forced runs}

In the forced runs (main text Fig. 2), our active dimer fluid is contained in a periodic box of side length $L = 100 \, \sigma$ at a density $\rho_0 \approx 0.41 \, m \, \sigma^{-2}$, matching the bulk density obtained in the boundary current simulations. We subject each monomer to a sinusoidal force field
\begin{equation}
{\boldsymbol F} \left( y \right) = F_0 {\rm sin} \left( k y \right) \hat{\boldsymbol x}
\label{eqn:force_field_monomer}
\end{equation}
with $k \equiv 2 \pi n / L$ and $n = 5$. This corresponds to a force per unit volume on each dimer
\begin{equation}
\begin{split}
{\boldsymbol f} \left( y_i^1, y_i^2 \right) &= \frac{\rho_0}{2 m} \sum_{\alpha} {\boldsymbol F} \left( y_i^{\alpha} \right) \\
&= \frac{\rho_0 F_0}{2 m} \left[ {\rm sin} \left( k y_i^1 \right) + {\rm sin} \left( k y_i^2 \right) \right]
\end{split}
\label{eqn:force_field_hydrodynamic_initial}
\end{equation}
The sum of sines may be expanded to second lowest order in $k {\boldsymbol d}_i \cdot \hat{\boldsymbol y}$ and averaged orientationally to give
\begin{equation}
\begin{split}
&{\rm sin} \left( k y_i^1 \right) + {\rm sin} \left( k y_i^2 \right) \\
&= 2 {\rm sin} \left( k y_i \right) - \frac{\left( k d_0 \right)^2}{4} \left\langle \left( \hat{\boldsymbol d}_i \cdot \hat{\boldsymbol y} \right)^2 \right\rangle_{\theta,y} {\rm sin} \left( k y_i \right) \\
& \quad \quad \quad \quad \quad \quad \quad \quad+ \mathcal{O} \left( \left( kd_0 \right)^4 \right)
\end{split}
\label{eqn:sine_expansion}
\end{equation}
where the subscript $\theta,y$ on the angled brackets indicates that it is an orientational overage that may in principle depend on the location $y$. However, if linear response is valid, the magnitude of the applied force $F_0$ must be weak enough that the force field does not locally modify the orientational structure of the fluid, and the orientational average is therefore unbiased and independent of location. We have confirmed that this is the case by conducting simulations with different values of $F_0$ and ensuring that the results are independent of this parameter. We thus have $\langle (\hat{\boldsymbol d}_i \cdot \hat{\boldsymbol y})^2 \rangle = 1/2$, and Eq. \ref{eqn:sine_expansion} becomes
\begin{equation}
\begin{split}
&{\rm sin} \left( k y_i^1 \right) + {\rm sin} \left( k y_i^2 \right) \\
&= 2 {\rm sin} \left( k y_i \right) - \frac{\left( k d_0 \right)^2}{8} {\rm sin} \left( k y_i \right) + \mathcal{O} \left( \left( kd_0 \right)^4 \right)
\end{split}
\label{eqn:sine_expansion2}
\end{equation}
$\left( k d_0 \right)^2/8 \sim \mathcal{O} \left( 0.01 \right)$ for our simulation parameters, and we therefore conclude that the second term is negligible. We thus have for the hydrodynamic body force per unit volume
\begin{equation}
{\boldsymbol f} \left( y \right) = \frac{\rho_0 F_0}{m} {\rm sin} \left( k y \right)
\label{eqn:hydrodynamic_force}
\end{equation}

We solve Eqs. \ref{eqn:continuity_incompressible}, \ref{eqn:momentum_incompressible}, and \ref{eqn:Deltaspin_incompressible} for a constant density fluid with $\phi = \alpha_e = 0$, consistent with the dynamic regime observed for the boundary current, and subjected to the body force ${\boldsymbol f} \left( y \right)$ defined in Eq. \ref{eqn:hydrodynamic_force}. We anticipate that solutions must be of the form
\begin{equation}
{\boldsymbol u} = \hat{u} \, {\rm sin} \left( k y \right) \hat{\boldsymbol x}
\label{eqn:forced_velocity_solution_form}
\end{equation}
\begin{equation}
p = \hat{p} \, {\rm cos} \left( k y \right)
\label{eqn:forced_pressure_solution_form}
\end{equation}
and
\begin{equation}
\Delta \mathcal{L} = \hat{\mathcal{L}} \, {\rm cos} \left( k y \right)
\label{eqn:forced_velocity_solution_form}
\end{equation}
This form of the velocity field automatically satisfies the solenoidal condition, Eq. \ref{eqn:continuity_incompressible}, and the Fourier amplitudes of the velocity and spin are therefore from Eqs. \ref{eqn:momentum_incompressible} and \ref{eqn:Deltaspin_incompressible} related to the amplitude $\rho_0 F_0 / m$ of the force field (Eq. \ref{eqn:hydrodynamic_force}) by
\begin{equation}
\hat{u} = \frac{1}{1 + \left( k / \kappa_{\gamma} \right)^2} \frac{F_0}{\gamma}
\label{eqn:u_forced}
\end{equation}
\begin{equation}
\hat{p} = - \frac{k/\kappa_{\gamma}}{1 + \left( k / \kappa_{\gamma} \right)^2} \frac{\eta_O \kappa_{\gamma} F_0}{\gamma}
\label{eqn:p_forced}
\end{equation}
and
\begin{equation}
\hat{\mathcal{L}} = - 2 \mathcal{N}_R \frac{k/\kappa_{\gamma}}{1 + \left( k / \kappa_{\gamma} \right)^2} \frac{\kappa_{\gamma}^{-1} F_0}{\gamma}
\label{eqn:L_forced}
\end{equation}
with $\kappa_{\gamma}$ defined in Eq. \ref{eqn:friction_lengths} and $\mathcal{N}_R$ defined in Eq. \ref{eqn:dim_params}. We see that the forms obtained for the velocity and spin, Eqs. \ref{eqn:u_forced} and \ref{eqn:L_forced}, allow us to test directly the values of $\eta_S$ and $\eta_R$ obtained in the boundary current simulations against the results of our forced simulations (main text Fig. 2).
